\documentclass{jfm}

\usepackage{graphicx}
\usepackage{xcolor}
\usepackage{upgreek}
\usepackage{epstopdf,epsfig}
\usepackage{newtxtext}
\usepackage{newtxmath}
\usepackage{natbib}
\usepackage{hyperref}
\hypersetup{
    colorlinks = true,
    urlcolor   = blue,
    citecolor  = black,
}

\newcommand{\RomanNumeralCaps}[1]
\linenumbers

\title{Plane-marching PSE wavepacket models for perfectly-expanded twin jets}

\author{Iv\'{a}n Padilla-Montero\aff{1}
  \corresp{\email{ivan.padilla@upm.es}},
  Daniel Rodr\'{i}guez\aff{1},
  Vincent Jaunet\aff{2}
 \and Peter Jordan\aff{2}}

\affiliation{\aff{1}School of Aeronautics (ETSIAE), Universidad Polit\'{e}cnica de Madrid, 28040 Madrid, Spain
\aff{2}D\'{e}partement Fluides, Thermique et Combustion, Institut Pprime, CNRS - Universit\'{e} de Poitiers - ISAE-ENSMA, 86036 Poitiers, France}

\begin{document}
\maketitle

\begin{abstract}
The importance of wavepackets in the generation of mixing noise in twin jets is expected by extrapolation of the insights previously obtained from the study of single isolated jets. This work presents wavepacket models for supersonic round twin jets operating at perfectly-expanded conditions, computed via plane-marching parabolized stability equations based on mean flows obtained from the compressible RANS equations. High-speed schlieren visualizations and non-time-resolved PIV measurements are performed to obtain experimental datasets for validating the modelling strategy. The RANS solutions are found to be in good quantitative agreement with the PIV mean-flow measurements, confirming the ability of the approach to capture the interaction between jets at the mean-flow level. The obtained wavepackets consist of toroidal and flapping fluctuations of the twin-jet system, and show similarities with those of single axisymmetric jets. However, for the case of closely-spaced jets, they exhibit deviations in the phase speed of structures travelling in the outer mixing layer and those travelling in the inner one, leading to different non-axisymmetric behaviours. In particular, toroidal twin-jet wavepackets feature tilted ring-like structures with respect to the jet axis, while flapping twin-jet wavepackets are distorted and lose the clean checkerboard pattern typically observed in $m = 1$ modes in axisymmetric jets. A quantitative comparison of the modelled wavepackets with experimentally-educed coherent structures is performed in terms of their structural agreement measured through an alignment coefficient, providing a first validation of the modelling strategy. Alignment coefficients are found to be particularly high in the intermediate range of studied frequencies.
\end{abstract}


\section{Introduction}
\label{sec:intro}

Jet noise is an environmental challenge that the aerospace industry must face today and for years to come. High-speed aircraft and space launchers commonly feature multi-jet engines to supply the required levels of thrust, with twin-jet configurations being the most common. Twin-jet flow fields present a number of challenges in comparison to single jets, which have been widely studied in the past. When close to each other, the jets may interact at the hydrodynamic and acoustic levels, yielding complex physical mechanisms that remain to be understood. Early experiments studying twin-jet configurations~\citep{Bhat:AIAA77,Kantola:JSV81} identified mechanisms which result in a reduction of the far-field noise with respect to an equivalent isolated jet, due in part to an acoustic shielding of one jet by the other. On the contrary, recent experimental observations by~\citet{Bozak:AIAA2011} have revealed enhanced noise levels in the plane contained between the two jets, which are above those corresponding to two linearly superposed incoherent jets.

The development of jet-noise reduction and control technologies requires understanding and appropriate simplified models for the mechanisms of sound generation. This has motivated study of the dynamics of turbulent shear flow. In this regard, the importance of ordered motions (today known as coherent structures) in turbulent planar mixing layers and jets has been recognized since the pioneer experimental observations by~\citet{Mollo-Christensen1967}, \citet{CrowChampagne:JFM71} and~\citet{Brown:JFM1974}. Since then, the connection between the sound radiated by high-speed jets and the coherent structures has been the focus of significant research, as reviewed by~\citet{JordanColonius:ARFM13}. Coherent structures were found to be reminiscent of linear instability waves growing in harmonically-forced supersonic jets, a finding that prompted the use of linear stability theory for their modelling~\citep{CrowChampagne:JFM71,CrightonGaster:JFM76,MichalkePrAS1984,TamMorris:JSV1985} and established the term ``wavepacket'' to refer to them. Since those studies, the presence of wavepackets in unforced high-speed jets and their relationship with mixing-noise emission has been demonstrated both from experimental analyses~\citep{Juve:JSV80,Guj:JSV2003} and large-eddy simulations~\citep{Cavalieri:JSV11b}, together with the finding that the radiated sound is highly directional both for subsonic and perfectly-expanded supersonic jets~\citep{CrightonHuerreJFM90,Tam:ARFM95,Cavalieri:JFM12}. In parallel, several works consolidated the ability of linear stability calculations to model wavepackets correctly, both in subsonic~\citep{Suzuki:JFM06,Piot:IJAAF06,Gudmundsson:JFM11,Cavalieri:JFM13} as well as supersonic jets~\citep{Sinha:JFM14}. Although wavepackets constitute a relatively small fraction of the total fluctuation energy of the turbulent flow~\citep{Cavalieri:JFM13}, they nevertheless are known to be acoustically significant owing to their high spatio-temporal coherence with respect to the small turbulent scales~\citep{JordanColonius:ARFM13}.

Among the different linear stability theories employed for wavepacket modelling in single jets, the parabolized stability equations (PSE) have been broadly used. PSE were found to deliver a satisfactory agreement with both high-fidelity simulations and experiments for subsonic and supersonic jets at a small computational cost~\citep{YenMessersmithAIAA1999,Piot:IJAAF06,Ray:PF09,Gudmundsson:JFM11,Rodriguez:AIAA12,Cavalieri:JFM13,Rodriguez2013,Sinha:JFM14,Breakey:PRF2017,Sasaki:JFM17}. These works have also served to identify and address some limitations of the PSE model for jet noise, which may be overcome by more computationally expensive approaches such as the one-way Navier-Stokes equations~\citep{Towne:JCP15,Towne:JFM2022} or resolvent analysis~\citep{Garnaud:JFM2013,Jeun:PF2016,Schmidt:JFM2018,Cavalieri:ARM2019}. On the one hand, difficulties have been found in predicting the wavepacket properties at low frequencies and low azimuthal wavenumbers, conditions at which recent studies based on resolvent analysis have found non-modal growth to be important through the Orr and lift-up mechanisms~\citep{Tissot:PRF2017,Schmidt:JFM2018,Lesshafft:PRF2019,Nogueira:JFM2019,Pickering:JFM2020}. Similarly, discrepancies have been encountered in the PSE description of wavepackets downstream of the region at which the growth of Kelvin-Helmholtz instabilities saturates, where locally-parallel transient growth calculations and resolvent analyses reveal non-modal phenomena to be important and point to the role of non-linear interactions in their activation~\citep{Tissot:PRF2017,Jordan:AIAAC2017,Schmidt:JFM2018,Lesshafft:PRF2019}. On the other hand, an underprediction of the amplitude of Mach-wave radiation has been observed, which can be attributed to the inherent limitations of the regularized marching problem, as described by~\citet{Towne:TCFD2019}. Despite these shortcomings, coherent structures in supersonic jets at the frequency range of interest in the study of mixing noise are dominated by Kelvin-Helmholtz instabilities~\citep{Pickering:JFM2020}, which are well modelled by PSE owing to their modal convective nature. For supersonic mixing layers featuring Kelvin-Helmholtz waves with high (supersonic) phase speed, the error in the PSE representation of Mach-wave radiation is found to be small~\citep{Towne:TCFD2019}. These conditions are satisfied for most wavepacket calculations considered in this work.

Compared to single round jets, studies of wavepackets and their modelling in twin-jet systems are scarce in the literature, mainly due to the increased complexity of the flow field. The twin-jet mean flow is no longer axisymmetric, which prevents the introduction of azimuthal Fourier modes in both the modelling approaches and the experimental postprocessing techniques, and in turn requires inhomogeneity in at least two spatial directions. From the modelling point of view,~\citet{Sedelnikov1967} was the first to derive dispersion relationships for multi-jet configurations employing vortex sheets, although no solutions were determined. After this, only three works addressed the parallel-flow linear instability problem for twin jets before the last decade~\citep{Morris:JFM90,Du1993,GreenCrighton:JFM97}, making use of vortex sheet and finite-thickness models based on bipolar coordinate systems to study the inviscid instability of two-axially homogeneous parallel jets. \citet{Morris:JFM90} established the classification of twin-jet fluctuations into four possible families according to the natural symmetries of the system. More recently, cross-plane local linear stability theory~\citep{Rodriguez2018,Nogueira2021,Rodriguez2023JFM}, along with plane-marching parabolized stability equations (PM-PSE)~\citep{Rodriguez2018,Rodriguez2021} have enabled characterization of the three-dimensional structure of the Kelvin-Helmholtz instabilities associated with mixing noise in the round twin-jet system. \citet{Rodriguez2018} computed wavepacket models for subsonic twin jets employing local stability analyses and plane-marching PSE on a tailored twin-jet mean flow, consisting of the linear superposition of two single-jet analytical velocity fields fitted from hot-wire measurements. Similarly,~\citet{Rodriguez2021} obtained wavepacket models for perfectly-expanded supersonic twin jets via PM-PSE, once again using a tailored twin-jet mean flow fitted from LES calculations. \citet{Nogueira2021} characterized Kelvin-Helmholtz instabilities in a supersonic twin-jet system operating at underexpanded conditions via a locally-parallel Floquet stability analysis, featuring a twin-jet mean flow obtained from planar PIV measurements revolved around the nozzle axes. Last, \citet{Rodriguez2023JFM} revisited the locally-parallel stability of perfectly-expanded twin jets using a vortex sheet and a finite-thickness model based on a tailored analytical mean flow, revealing that the coupling between the fluctuation fields of the two jets favours flapping motions over helical ones. These works employed simplified mean-flow models that account for linear interaction between the two jets. However, experimental mean-flow measurements and high-fidelity simulations indicate that the interaction is non-linear, particularly in the case of closely-spaced jets.

The aforementioned modelling efforts have predicted wavepackets analogous to those in single round jets, although with observable deviations in their spatial structure from the azimuthal Fourier modes. The few experimental works aimed at their identification and characterization in round twin-jet systems are very recent and focused on conditions at which screech is dominant~\citep{Kuo:AIAAJ2017,Knast:AIAAJ2018,Bell2021,Nogueira2021,Nogueira2021b,Wong2023}. Therefore, while the previously described investigations have provided physically-sound wavepacket models for mixing noise, an assessment of their validity is still missing. Significant experimental work has also focused on the study of the coupling and control of twin rectangular jets under screech conditions \citep{Raman:IJA2012,Esfahani2021,Yeung:AIAAC2022,Samimy:JFM2023,Jeun:JFM2024,Karnam:JFM2025}. However, no quantitative confrontation between wavepacket models and experimental measurements currently exist for such rectangular configurations either.

This work contributes to the modelling of wavepackets associated with mixing noise in supersonic round twin jets in two new ways: (i) by employing a more accurate mean-flow representation based on the compressible RANS equations, which is used as an input for the plane-marching parabolized stability equations, and (ii) by performing the first quantitative comparisons of twin-jet wavepackets against coherent structures educed from experimental measurements. The use of a three-dimensional RANS mean flow allows the non-linear interaction between both jets at a mean-flow level, following which wavepacket predictions can be made using plane-marching PSE. Here, RANS solutions are computed for perfectly-expanded twin jets and validated by means of particle image velocimetry (PIV) measurements of the mean flow.

The validation of twin-jet wavepackets computed by PSE against experimental data is realized by application of spectral proper orthogonal decomposition (SPOD) to high-speed schlieren visualizations. Recent investigations focused on the study of screech in single~\citep{Edgington-Mitchell2022,Karnam2023} and twin jets~\citep{Esfahani2021,Nogueira2021b,Prasad2022b,Wong2023,Karnam:JFM2025} have educed organised structures at resonant frequencies by means of SPOD applied to schlieren measurements. In this case, the dynamics of the resonant mechanism feature a strong low-rank behaviour and a highly-organized structure that is contained in the most energetic SPOD mode. The eduction of coherent structures from schlieren visualizations in perfectly-expanded twin-jet systems is more challenging, as the schlieren images under non-screeching conditions are dominated by highly energetic, high-azimuthal wavenumber, small-scale vortical structures localized in the turbulent shear layer~\citep{Cavalieri:JFM13}. These structures dominate the fluctuation energy of the turbulence and complicate the use of techniques like proper orthogonal decomposition in obtaining a low-rank representation of wavepackets.

In this work, recent improvements~\citep{Prasad2022,PadillaMontero2024} that facilitate the extraction of mixing-noise-related coherent structures from schlieren images are adopted. These consist in performing spectral proper orthogonal decomposition on a filtered quantity derived from the schlieren images, instead of applying it to the schlieren fields directly. The quantity used is intimately related to the irrotational momentum potential field introduced by~\citet{Doak1989}, which does not feature small-scale vortical fluctuations present in the schlieren visualizations. Through this methodology, coherent structures are extracted from the experimental datasets in the form of SPOD modes which can be compared with the PM-PSE wavepackets. To quantify the agreement between both, an alignment metric based on the projection of one SPOD field into one PM-PSE fluctuation field is proposed and evaluated for different frequencies and jet separations. Due to the line-of-sight integration implicit in the schlieren visualizations, only those fluctuations which are symmetric with respect to the plane containing both jets can be realized from the employed experimental setup. The analysis and validation of wavepackets which are antisymmetric with respect to this plane are thus not considered in this work.

The remainder of this paper is organized as follows. Section~\ref{sec:twinjet_config} describes the twin-jet geometry and conditions considered for study. Section~\ref{sec:modelling} presents the adopted modelling strategy for twin-jet wavepackets, formulating the governing equations as well as providing details on the numerical methodology employed for the computations. In section~\ref{sec:exp_setup}, the twin-jet experimental setup is reported, including a description of the method employed for extracting coherent structures from the experimental data. Section~\ref{sec:results} presents the obtained mean-flow solutions and wavepacket models. Particular characteristics of the twin-jet fluctuations for closely-spaced jets are discussed, and their comparison against the experimentally-educed structures is reviewed. Finally, concluding remarks are summarized in section~\ref{sec:conclusions}.

\section{Twin-jet configuration}
\label{sec:twinjet_config}

\begin{figure}
\centerline{\includegraphics[width=0.85\textwidth]{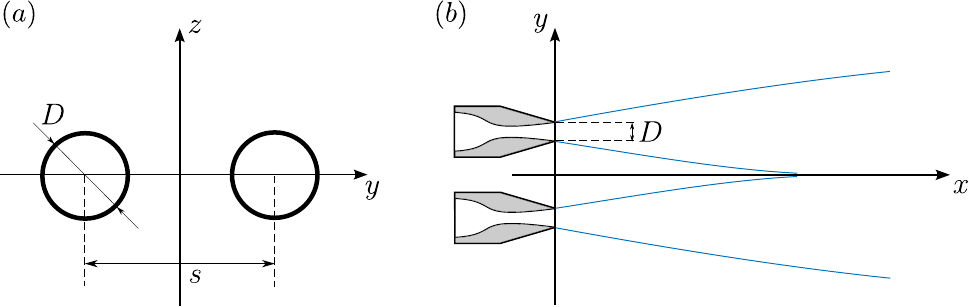}}
\caption{Sketch representing the twin-jet configuration and the associated geometrical parameters: $(a)$ cross-stream plane (constant $x$); $(b)$ streamwise symmetry plane containing both jets at $z = 0$.}
\label{fig:twinjet_geometry}
\end{figure}

The twin-jet configuration considered is sketched in figure~\ref{fig:twinjet_geometry}. The jets are generated by two round convergent-divergent nozzles. The center of each of the nozzles is located along the $y$-axis, with $y = 0$ denoting the symmetry plane between the jets. Each nozzle exit is located at $x = 0$, and the spacing between the nozzles axes is denoted by $s$. The interior nozzle geometry follows a truncated ideal contour (TIC) profile with an exit diameter of $D = 0.025$ m and an exit-to-throat area ratio of $A_e / A_t = 1.225$. The nozzle geometry is shown in figure~\ref{fig:RANS_single_jet}.

The jets are generated under a fixed nozzle pressure ratio (NPR) corresponding to the perfectly-expanded condition and a fixed total temperature of $T_{0} = 300$ K. The nozzle pressure ratio, defined as the ratio of the total pressure in the reservoir $p_{0}$ to the ambient pressure $p_\infty$, is expressed in terms of the isentropic jet-exit Mach number $M_j$ through the isentropic relation $p_{0} / p_\infty = [1 + 0.5(\gamma - 1) M_j^2 ]^{\gamma/(\gamma - 1)}$, where $\gamma = 1.4$ denotes the ratio of specific heats. The jets are unheated, and the flow acceleration within the nozzles results in jet static temperatures lower than the ambient temperature. The operating condition closest to the perfectly-expanded regime has been calibrated experimentally by means of flow visualization and corresponds to $M_j = 1.54$ (NPR = 3.89). This operating condition is considered throughout the study.

Two different nozzle separations are investigated: $s/D = 1.76$ and $s/D = 3$. The first corresponds to the minimum distance allowed between the nozzles axes due to the size of the outer nozzle diameter. A strong interaction between jets is expected for $s/D = 1.76$, while a weak interaction is expected for $s/D = 3$. The system exhibits two symmetry planes: the plane that contains both jets (the $xy$ plane located at $z = 0$) and the plane between the jets (the $xz$ plane located at $y = 0$).

\section{Modelling strategy}
\label{sec:modelling}

\subsection{Mean flow}

In the past, twin-jet mean-flow descriptions have been obtained using large eddy simulations~\citep{Bres2014,Gao2016,Goparaju2018,Bres:AIAAC2021,Troyes2022,Muthichur:JFM2023} or simplified approaches such as the axisymmetric revolution of twin-jet planar PIV measurements~\citep{Nogueira2021} or via a tailored twin-jet model, in which the twin-jet flow field is constructed by linearly superposing the velocity fields of two single jets. In the later, the single-jet mean fields may come either from analytical models such as hyperbolic tangent profiles~\citep{Stavropoulos2023,Rodriguez2023JFM} or analytical gaussian fittings to single-jet PIV measurements~\citep{Gudmundsson:JFM11}, hot-wire measurements~\citep{Rodriguez2018} or LES calculations~\citep{Rodriguez2021}. 

One of the goals of this work is to model wavepackets in supersonic twin-jets by means of linear parabolized stability equations based on an accurate mean-flow representation. A mean-flow description that properly accounts for the non-linear interactions between the jets at a mean-flow level is thus preferred. Three-dimensional compressible RANS calculations are used to capture the mean-flow jet interaction with a good accuracy at a reduced computational cost compared to LES.

\subsubsection{Governing equations}

The twin-jet mean flow is modelled using the compressible RANS equations (also known as Favre-averaged Navier-Stokes equations), see e.g.~\citet{Wilcox2006}:

\begin{equation}
\frac{\partial \bar{\rho}}{\partial t} + \frac{\partial}{\partial x_i} \left( \bar{\rho} \tilde{u}_i \right) = 0,
\end{equation}

\begin{equation}
\frac{\partial \left( \bar{\rho} \tilde{u}_i \right)}{\partial t} + \frac{\partial}{\partial x_j} \left( \bar{\rho} \tilde{u}_i \tilde{u}_j \right) = -\frac{\partial \bar{p}}{\partial x_i} + \frac{\partial}{\partial x_j} \left( \bar{\tau}_{ij} - \overline{\rho u_i'' u_j''} \right),
\end{equation}

\begin{multline}
\frac{\partial}{\partial t} \left( \bar{\rho} \tilde{E} + \frac{\overline{\rho u_i'' u_i''}}{2} \right) + \frac{\partial}{\partial x_j} \left( \bar{\rho} \tilde{u}_j \tilde{H} + \tilde{u}_j \frac{\overline{\rho u_i'' u_i''}}{2} \right) = \frac{\partial}{\partial x_j} \left[ \tilde{u}_i \left( \bar{\tau}_{ij} - \overline{\rho u_i'' u_j''} \right) \right]\\
+ \frac{\partial}{\partial x_j} \left( \overline{\tau_{ij} u_i''} - \frac{\overline{\rho u_i'' u_i'' u_j''}}{2} \right) - \frac{\partial
}{\partial x_j} \left(-\bar{\kappa} \frac{\partial \tilde{T}}{\partial x_j} + c_p \overline{\rho T'' u_j''} \right),
\end{multline}

\noindent where $u_i$ is the velocity component along the $i$th direction (with $i = 1,2,3$), $\rho$ is the density, $p$ denotes pressure and $T$ the temperature. For a given instantaneous flow variable $q$, the following decompositions are applied: $q = \bar{q} + q' = \tilde{q} + q''$. The symbol $\bar{\cdot}$  indicates conventional time averaging, whereas $\tilde{\cdot}$ denotes Favre averaging, defined as

\begin{equation}
\tilde{q} = \frac{\overline{\rho q}}{\bar{\rho}}.
\end{equation}

The quantity $\bar{\tau}_{ij}$ is the viscous stress tensor, given by

\begin{equation}
\bar{\tau}_{ij} = \bar{\mu} \left( \frac{\partial \tilde{u}_i}{\partial x_j} + \frac{\partial \tilde{u}_j}{\partial x_i} \right) - \frac{2}{3} \bar{\mu} \frac{\partial \tilde{u}_k}{\partial x_k} \delta_{ij},
\end{equation}

\noindent where $\delta_{ij}$ is the Kronecker delta. The quantity $-\overline{\rho u_i'' u_j''}$ denotes the Favre-averaged Reynolds stress tensor, which is modelled by means of the Boussinesq eddy-viscosity approximation. $\tilde{E}$ refers to the Favre-averaged total energy, defined as

\begin{equation}
\tilde{E} = c_v \tilde{T} + \frac{\tilde{u}_i \tilde{u}_i}{2},
\end{equation}

\noindent where $c_v$ represents the specific heat at constant volume, while $\tilde{H}$ is the Favre-averaged total enthalpy, given by $\tilde{H} = \tilde{E} + \bar{p}/\bar{\rho}$. The quantity $\frac{1}{2} \overline{\rho u_i'' u_i''} = \bar{\rho} k$ denotes the kinetic energy of the turbulent fluctuations per unit volume. The transport coefficients representing the dynamic viscosity and thermal conductivity are respectively denoted by $\bar{\mu}$ and $\bar{\kappa}$. The evolution of viscosity with temperature follows Sutherland's law and $\bar{\kappa}$ is obtained from the assumption of a constant Prandtl number. Finally, the averaged perfect gas equation of state reads

\begin{equation}
\bar{p} = \bar{\rho} R_g \tilde{T},
\end{equation}

\noindent where $R_g = 287$ J/(kg K) is the specific gas constant.

\begin{table}
  \begin{center}
\def~{\hphantom{0}}
  \begin{tabular}{ccccccccc}
      $a$ & $\beta^*$ & $\kappa$ & $\beta_1$ & $\beta_2$ & $\sigma_{k1}$ & $\sigma_{k2}$ & $\sigma_{\omega 1}$ & $\sigma_{\omega 2}$ \\[3pt]
      0.372 & 0.085 & 0.41 & 0.0726 & 0.0924 & 0.662 & 0.787 & 0.446 & 0.745\\
  \end{tabular}
  \caption{Employed values for the parameters of the Menter SST turbulence model. The same nomenclature as in \citet{Menter1994} is followed.}
  \label{tab:SST_constants}
  \end{center}
\end{table}

The two-equation SST turbulence model introduced by~\citet{Menter1994} is considered, together with the compressible mixing-layer correction proposed by~\citet[eq. 5.83]{Wilcox2006}. Modified values of the empirical constants of the turbulence model have been employed and are listed in table~\ref{tab:SST_constants}. These values were obtained by manual calibration following a linear interpolation between the original SST constants provided in~\citet{Menter1994} and the optimized constants recently obtained by~\citet{Ozawa2024} via data assimilation for a perfectly-expanded jet. The selected values were found to yield a small mean squared error with respect to PIV mean-flow measurements with the employed solver. Appendix~\ref{appA} provides a comparison between the mean-flow results obtained with the three different sets of empirical constants, which justifies the choice of parameters for this work.

The reference quantities employed for non-dimensionalization are the nozzle-exit diameter $D$ and the freestream (ambient) flow conditions. The flow velocity components are non-dimensionalized with the freestream speed of sound $c_\infty$; the density with the respective freestream value $\rho_\infty$; the pressure is made dimensionless with $\rho_\infty c_\infty ^2$ and the temperature with $(\gamma - 1) T_\infty$. The transport properties $\mu$ and $\kappa$ are non-dimensionalized with their respective freestream values.

\subsubsection{Numerical methodology for the RANS calculations}

\begin{figure}
\centering
\includegraphics[width=0.88\textwidth]{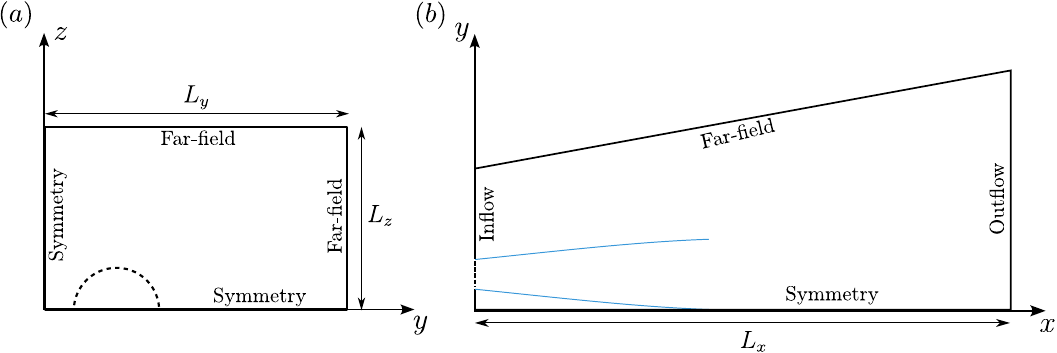}
\caption{Computational domain employed for the RANS calculation of the twin-jet flow field, including boundary conditions: $(a)$ cross-stream plane at a fixed streamwise location; $(b)$ $xy$ symmetry plane plane at $z = 0$. The dashed lines mark the nozzle exit geometry.}
\label{fig:TJ_comp_domain}
\end{figure}

\begin{table}
  \begin{center}
\def~{\hphantom{0}}
  \begin{tabular}{ccccccccc}
      $p_0$ [Pa] & $T_0$ [K] & $p_{\infty}$ [Pa] & $T_\infty$ [K] & $M$ & $M_j$ & $\Rey$ & $\Rey_j$ & $\Pran$ \\[3pt]
      $3.891 \times 10^5$ & $300$ & $10^5$ & $300$ & $1.27$ & $1.54$ & $5.460 \times 10^5$ & $1.398 \times 10^6$ & $0.72$\\
  \end{tabular}
  \caption{Flow conditions employed for the RANS calculations and resulting dimensionless parameters.}
  \label{tab:flow_conditions}
  \end{center}
\end{table}

\begin{figure}
\centerline{\includegraphics[width=0.99\textwidth]{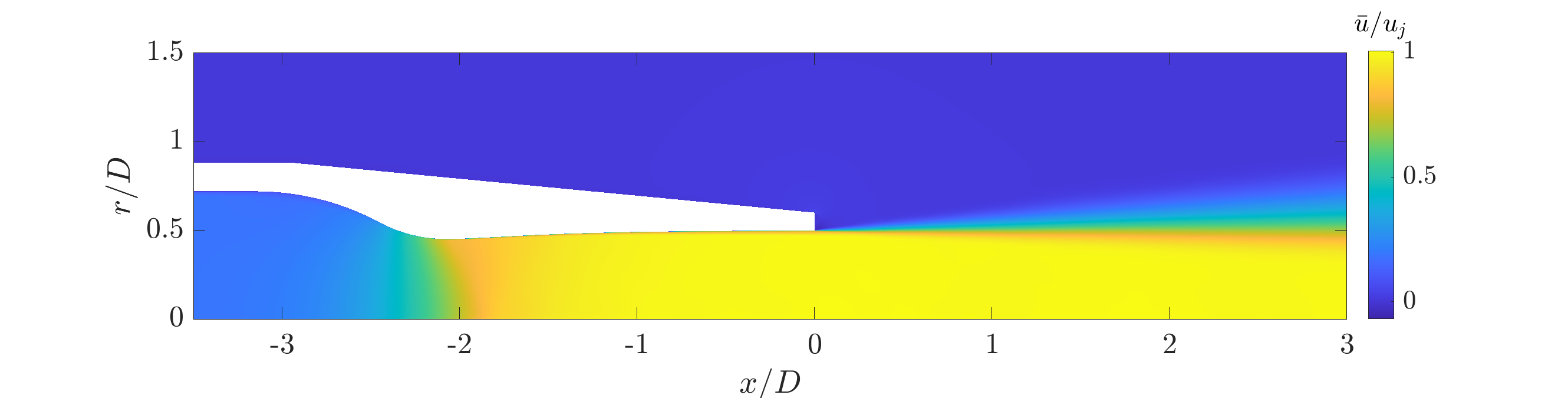}}
\caption{Contours of mean streamwise velocity for the axisymmetric (single) jet calculation.}
\label{fig:RANS_single_jet}
\end{figure}

\begin{figure}
\centerline{\includegraphics[width=0.99\textwidth]{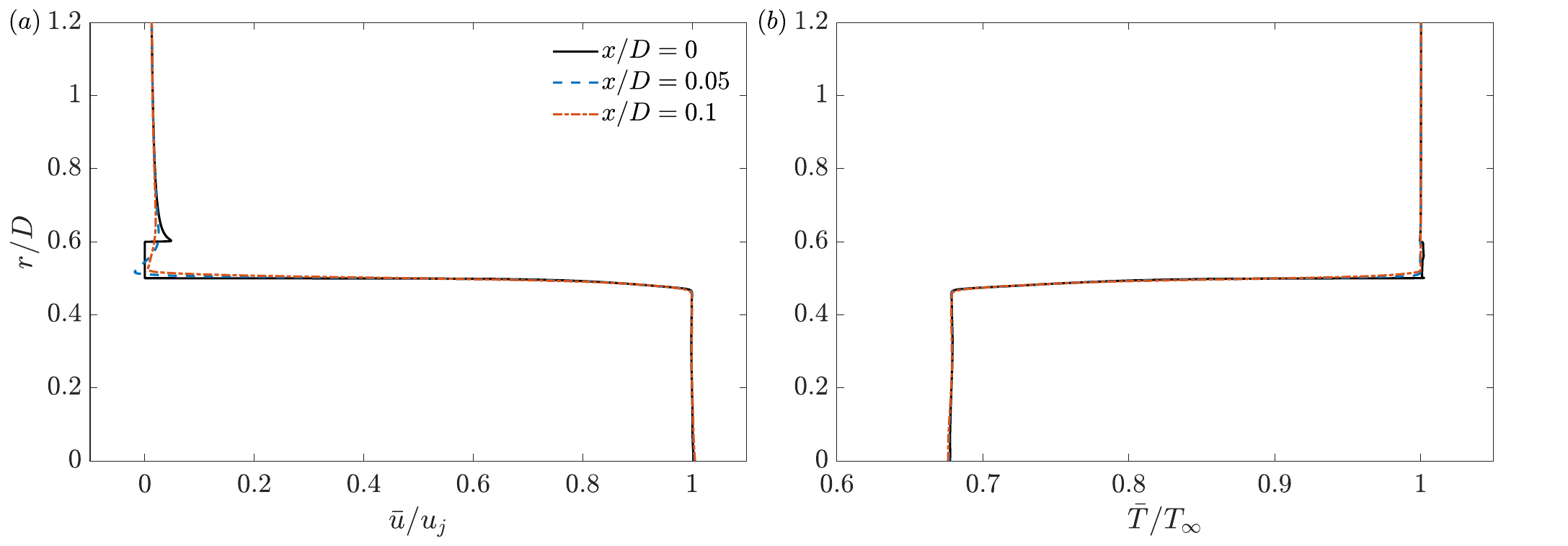}}
\caption{Mean flow profiles extracted at the nozzle exit ($x/D = 0$) and near the nozzle exit ($x/D = 0.05$ and $x/D = 0.1$) from the axisymmetric (single jet) calculation: $(a)$ streamwise velocity; $(b)$ static temperature.}
\label{fig:RANS_nzz_exit_profile}
\end{figure}

The RANS calculations are performed using the CFD solver TAU, developed by the German Aerospace Center (DLR)~\citep{Schwamborn2006}. The numerical solution employs a second-order finite-volume method with an upwind discretization scheme based on Roe's approximate Riemann solver. A steady RANS solution is targeted, which is approached using a backward Euler implicit scheme for time integration.

Figure~\ref{fig:TJ_comp_domain} shows the computational domain employed for the three-dimensional RANS calculations. Only one quarter of the twin-jet system is simulated, taking advantage of the two symmetry planes inherent in the geometry. The size of the domain is $L_x = 35D$ in the streamwise direction. In the $z$ direction, a size of $L_z = 8D$ is used at the domain inflow, which extends linearly until $L_z = 14D$ at the domain outflow. Similarly, in the $y$ direction, $L_y = 9D$ is employed at the inflow, while $L_y = 15D$ is used at the outflow. The inflow is located at the nozzle exit plane ($x/D = 0$), where the nozzle exit flow field is imposed as a Dirichlet boundary condition. The flow at the nozzle exit is computed by means of an axisymmetric compressible RANS calculation of the entire nozzle flow field, including the boundary layer developing on the nozzle inner walls. This solution is shown in figure~\ref{fig:RANS_single_jet}, which shows contours of the streamwise velocity field inside the nozzle and in the vicinity of the nozzle exit, and in figure~\ref{fig:RANS_nzz_exit_profile}, which illustrates the mean velocity and temperature profiles at the nozzle exit ($x/D = 0$) and slightly downstream of it ($x/D = 0.05$ and $0.1$). The profile corresponding to $x/D = 0$ is the one imposed at the inflow of the twin-jet calculation.

The single-jet axisymmetric RANS calculation uses a two-dimensional computational domain with a hybrid mesh mainly consisting of triangular cells, together with a structured region of quadrilateral cells located at the nozzle boundary layer. The cell size is calibrated such that $y^+ \approx 1$ to resolve the boundary layer developing on the nozzle walls. A small co-flow of $M_\textrm{co} = 0.01$ is included in the far-field to replicate the experimental conditions. Table~\ref{tab:flow_conditions} reports the total and the ambient flow conditions employed for both the single-jet and the twin-jet RANS calculations. Besides the jet-exit Mach number $M_j = u_j/c_j$, the acoustic jet Mach number is also reported, defined as $M = u_j/c_\infty$. Here, $u_j$ is the jet-exit velocity and $c_j$ is the speed of sound corresponding to the jet-exit temperature ($T_j$), which is determined assuming an isentropic expansion. Two different Reynolds numbers are considered, the reference, acoustic Reynolds number used for non-dimensionalization, $\Rey = \rho_\infty c_\infty D / \mu_\infty$, and the jet-exit Reynolds number, $\Rey_j = \rho_j u_j D / \mu_j$, where $\rho_j$ and and $\mu_j$ respectively denote the density and the dynamic viscosity of the jet at the nozzle exit, which are also determined based on the isentropic flow relations. The Prandtl number is defined as $\Pran = c_p \mu_\infty / \kappa_\infty$, were $c_p$ is the specific heat at constant pressure.

For the three-dimensional calculations, a fully unstructured tetrahedral mesh is used. It comprises successive refinement regions downstream of the nozzle exit following the growth of the mixing layers. A total of 10 million tetrahedral cells are employed for the case $s/D = 1.76$ and 11 million cells for $s/D = 3$. At the symmetry planes $z = 0$ and $y = 0$ of the domain, symmetry boundary conditions are imposed. At the domain outflow, a subsonic outflow boundary condition is used, which imposes the ambient pressure $p_\infty$ and extrapolates the other flow variables from the interior of the domain. At the far-field boundaries, a far-field inflow/outflow boundary condition is imposed that evaluates boundary fluxes via a characteristics method.

\subsection{Formulation of the plane-marching parabolized stability equations}

The parabolized stability equations (PSE) constitute one approach for studying the growth of linear and non-linear disturbances in shear flows, and are well-adapted when there exists a slow divergence of the mean-flow properties in the streamwise direction. They have been shown to deliver results comparable to direct numerical simulations for convectively unstable laminar and transitional flows~\citep{Bertolotti:JFM92,ChangMalikErlenbacherHussaini93}. PSE has been extensively used as a model for the spatial evolution of coherent structures in single round turbulent jets (wavepackets), yielding good agreement with coherent structures educed either from experiments or large eddy simulations~\citep{YenMessersmithAIAA1999,Piot:IJAAF06,Gudmundsson:JFM11,Cavalieri:JFM13,Rodriguez2013,Sinha:JFM14,Sasaki:JFM17}. For perfectly-expanded supersonic jets, screech does not occur and the sound generated at this condition is largely dominated by downstream-propagating Kelvin-Helmholtz waves. Owing to its lower computational cost and complexity compared with alternative techniques, PSE remains an attractive simplified modelling framework for this problem.

Following recent investigations~\citep{Rodriguez2018,Rodriguez2021}, wavepackets associated with mixing noise in perfectly-expanded supersonic twin jets are modelled here using linear plane-marching parabolized stability equations (PM-PSE). This model is a direct extension of the classical PSE approach that accounts for mean flows featuring a strong inhomogeneity in the cross-stream planes, i.e., $y$ and $z$~\citep{Broadhurst:2008}.

The plane-marching parabolized stability equations employed in this study are derived from the compressible Navier-Stokes equations. To formulate the PM-PSE linear stability problem, first the instantaneous turbulent flow field is separated into a stationary mean-flow component $\bar{\mathbf{q}}$ and an unsteady fluctuation component $\mathbf{q}'$:

\begin{equation}
\mathbf{q} (x,y,z,t) = \bar{\mathbf{q}} (x,y,z) + \mathbf{q}'(x,y,z,t),
\end{equation}

\noindent where $\mathbf{q} = [u,v,w,p,T]^\mathrm{T}$ is the vector of primitive flow variables, with $u$, $v$ and $w$ representing the velocity component along $x$, $y$ and $z$, respectively. Next, the fluctuating component is expressed as a sum of discrete modes in frequency:

\begin{equation}
\mathbf{q}'(x,y,z,t) = \sum_\omega \hat{\mathbf{q}}_\omega (x,y,z) \mathrm{e}^{-\mathrm{i}\omega t} + c.c.,
\label{eq:PSE_ansatz}
\end{equation}

\noindent where $\omega = 2\upi f$ denotes the angular frequency and $c.c.$ stands for the complex conjugate. The Fourier-transformed fluctuations in time $\hat{\mathbf{q}}_\omega$ are then decomposed into a shape function ($\tilde{\mathbf{q}}_\omega$), which is assumed to undergo a slow variation along the streamwise direction $x$, and a wave function ($A_\omega$), which is allowed to vary rapidly with $x$:

\begin{equation}
\hat{\mathbf{q}}_\omega (x,y,z)  =  A_\omega(x) \ \tilde{\mathbf{q}}_\omega (x,y,z) 
 =  A_\omega(x_0) \exp\left(\mathrm{i} \int^x _{x_0} \alpha_\omega(\xi) \,\mathrm{d} \xi \right)\ \tilde{\mathbf{q}}_\omega (x,y,z),
\label{eq:PSE_ansatz_omega}
\end{equation}

\noindent where the complex quantity $\alpha_\omega = \alpha_r + \mathrm{i} \alpha_i$ is the streamwise wavenumber for frequency $\omega$, for which a slow variation with $x$ is also required. The streamwise coordinate $x_0$ is the location where the PSE integration is initialized. In the context of this work, this is typically a cross-section close to the nozzle exit. The quantity $A_\omega (x_0)$ refers to the initial disturbance amplitude, which is arbitrary in a linear problem. Introducing equations~\eqref{eq:PSE_ansatz} and \eqref{eq:PSE_ansatz_omega} into the linearized Navier-Stokes equations and performing an order of magnitude analysis to neglect terms of order $1/\Rey^2$, the following linear system is obtained:

\begin{equation}
\mathbf{L} \frac{\partial \tilde{\mathbf{q}}_\omega}{\partial x} = \mathbf{R} \tilde{\mathbf{q}}_{\omega},
\label{eq:PSE_system}
\end{equation}

\noindent where the linear operators $\mathbf{L}$ and $\mathbf{R}$ depend on the mean flow quantities and their first and second spatial derivatives, the complex spatial wavenumber $\alpha_\omega$ and real frequency $\omega$, and the non-dimensional parameters $\Rey$, $\Pran$ and $\gamma$.

The ansatz~\eqref{eq:PSE_ansatz_omega} allows the spatial disturbance growth to be accounted for by either the shape function or the wave function, making the solution non-unique. In order to solve this ambiguity, a normalization is imposed to eliminate the exponential dependence from $\tilde{\mathbf{q}}_\omega$, which takes the form~\citep{Herbert:ARFM97}:

\begin{equation}
\iint_\Omega \tilde{\mathbf{q}}^* _\omega \frac{\partial \tilde{\mathbf{q}}_\omega}{\partial x} \,\mathrm{d}y \mathrm{d}z = 0,
\label{eq:PSE_norm}
\end{equation}

\noindent where the superscript $^*$ denotes the complex conjugate and $\Omega$ is the cross-stream spatial domain in which the shape functions are defined for a given streamwise location.

In the following, the subscript $\omega$ will be dropped for convenience. Note that the quantities $\hat{\mathbf{q}}$, $\tilde{\mathbf{q}}$ and $\alpha$ are always associated with a given frequency.

\subsubsection{Initial condition: locally-parallel linear stability analysis}

The PM-PSE system takes the form of a downstream-marching problem in the streamwise coordinate $x$, which requires initial conditions at a given location $x_0$. In order to obtain a consistent set of initial conditions for the PM-PSE problem, a locally parallel stability problem is derived from the PSE formulation following~\citep{Rodriguez2013}. This is achieved by assuming that $\mathrm{d} \alpha / \mathrm{d} x = 0$, $\partial \hat{\mathbf{q}} / \partial x = \mathrm{i} \alpha \tilde{\mathbf{q}}$, and that, for a given streamwise station $x$, the mean flow is locally parallel: $\partial \bar{\mathbf{q}} / \partial x = 0$. These assumptions yield a generalized eigenvalue problem of the form:

\begin{equation}
\mathrm{i} \alpha \mathbf{L} \tilde{\mathbf{q}} = \mathbf{R} \tilde{\mathbf{q}}.
\label{eq:EVP}
\end{equation}

For a given cross-stream plane at an initial location $x_0$ and a given frequency $\omega$, the solution of the eigenvalue problem~\eqref{eq:EVP} delivers a set of complex eigenvalues and their corresponding two-dimensional eigenfunctions (defined on the cross-stream plane). For unbounded mean flows like the present ones, classical linear stability theory~\citep{Mack1984,Balakumar1992} shows the existence of a fixed number of continuous branches, related to the uniform flow surrounding the jets, in addition to an indefinite number of discrete eigenmodes associated with localized mean-flow shear. For a single round jet, the different eigenmode families are described by \citet{Rodriguez2013,Rodriguez2015}. Identification of the discrete modes of interest (Kelvin-Helmholtz instabilities in this case) from the local stability spectrum is achieved by visual inspection of the corresponding two-dimensional amplitude functions.

\subsubsection{Numerical methodology for the plane-marching PSE calculations}
\label{sec:PSE_numerics}

The numerical solution of both the marching problem and the local stability problem require the spatial discretization of the two-dimensional linear operators $\mathbf{L}$ and $\mathbf{R}$. Following the studies by~\citet{Gennaro:AIAAJ13} and~\citet{Rodriguez:ICOSAHOM16}, finite-difference stencils of seven points are used in both $y$ and $z$ directions for this purpose, offering a good trade-off between convergence of results and computational cost. According to the mean-flow symmetries, only a quarter of the domain needs to be discretized. Consequently, a rectangular domain $\Omega = [0,y_{\infty}]\times[0,z_{\infty}]$ is used for the cross-stream planes. Two independent coordinate transformations (mappings) are used to concentrate grid points in the jet mixing regions. The following transformation is used in both $y$ and $z$ directions:

\begin{equation}
\eta = \eta_c + (\eta_\infty - \eta_c) \frac{\sinh \left(a(\xi - b) \right)}{\sinh\left(a(1-b)\right)},
\label{eq:mapping}
\end{equation}

\noindent where $\eta$ is the mapped coordinate in the physical domain ($y$ or $z$), $\xi$ is the associated coordinate in the computational domain ($\xi \in [0,1]$); $\eta_c$ is the location of the jet axis, where the discretization is refined; $\eta_\infty$ is the maximum coordinate and $a$ is a real number that controls the intensity of the clustering of points. The remaining parameter $b$ must be determined for each combination of $\eta_c$, $\eta_\infty$ and $a$ so that $\eta = 0$ for $\xi = 0$. In the present calculations, a square domain $\Omega = [0,10] \times [0,10]$ is used, discretized with a spatial resolution of $N_y \times N_z = 201 \times 201$ points. For each case, the values $a = 5$, $y_c/D = s/(2D)$ and $z_c/D = 0.5$ are used. Mesh convergence tests, not reproduced here, ensure the numerical consistency of the results using these parameters.

The Cartesian coordinate system allows for the use of standard finite differences for the independent differentiation on $y$ and $z$, resulting in the differentiation matrices $\mathcal{D}_y$ and $\mathcal{D}_z$ for first-order derivatives and $\mathcal{D}_{yy}$ and $\mathcal{D}_{zz}$ for second-order derivatives. The same stencil is used for first- and second-order differentiation matrices, allowing for the control of the matrix structure required for efficiency of the sparse implementation. The cross-differentiation matrix is then obtained as $\mathcal{D}_{yz} = \mathcal{D}_y \times \mathcal{D}_z$.

The symmetry imposition is accomplished through appropriate sets of symmetry and antisymmetry boundary conditions on the perturbation quantities. Symmetry boundary conditions impose Neumann conditions on $\tilde{u}$, $\tilde{p}$ and $\tilde{T}$, Dirichlet on $\tilde{v}$ and Neumann on $\tilde{w}$ for the boundary at the $z$ axis, or Neumann on $\tilde{v}$ and Dirichlet on $\tilde{w}$ for the boundary at the $y$ axis. Antisymmetry boundary conditions impose Dirichlet conditions on $\tilde{u}$, $\tilde{p}$ and $\tilde{T}$, Neumann on $\tilde{v}$ and Dirichlet on $\tilde{w}$ for boundaries aligned with the $z$ axis, or Dirichlet on $\tilde{v}$ and Neumann on $\tilde{w}$ for boundaries aligned with the $y$ axis. 

For a given cross-stream plane, four possible sets of conditions can be imposed, leading to four solution families~\citep{Morris:JFM90,Rodriguez2018,Rodriguez2023JFM}: (i) symmetry conditions with respect to  both $z = 0$ and  $y = 0$ (denoted by SS); (ii) symmetry with respect to $z = 0$ and antisymmetry with respect to $y = 0$ (denoted by SA); (iii) antisymmetry with respect to $z = 0$ and symmetry with respect to $y = 0$ (denoted by AS); (iv) antisymmetry conditions with respect to both $z = 0$ and $y = 0$ (denoted by AA).

The local stability problem is solved by means of a parallelized sparse implementation of the shift-and-invert Arnoldi algorithm~\citep{Arnoldi1951}, which employs the MUMPS library~\citep{Amestoy2001} for solution of the linear systems of equations. An implicit Euler scheme is used to march the PM-PSE solution downstream. The resulting linear system of equations at each streamwise station is also solved via a parallel implementation using MUMPS. In order to adjust the value of $\alpha$ so that the normalization condition~\eqref{eq:PSE_norm} is satisfied, solution of the linear system is iterated together with the following relation:

\begin{equation}
\alpha^{(k+1)}_{j+1} = \alpha^{(k)}_{j+1} - \frac{\mathrm{i}}{\Delta x} \frac{\iint_{\Omega} \tilde{\textbf{q}}_{j+1}^* (\tilde{\textbf{q}}_{j+1} - \tilde{\textbf{q}}_{j}) \,\mathrm{d}y \mathrm{d}z}{\iint_\Omega \tilde{\textbf{q}}_{j+1}^* \tilde{\textbf{q}}_{j+1}  \,\mathrm{d}y \mathrm{d}z},
\end{equation}

\noindent where $k$ is the iteration index, $j+1$ refers to the solution at the next downstream streamwise position, $j$ refers to the current streamwise station and $\Delta x = x_{j+1} - x_j = 0.75D$ is the streamwise step size. The iteration continues until $\alpha$ is converged up to a relative error $|\alpha^{(k+1)}-\alpha^{(k)}|/|\alpha^{(k+1)}| < 10^{-4}$.

\section{Experimental setup}
\label{sec:exp_setup}

\begin{figure}
\centerline{\includegraphics[width=0.99\textwidth]{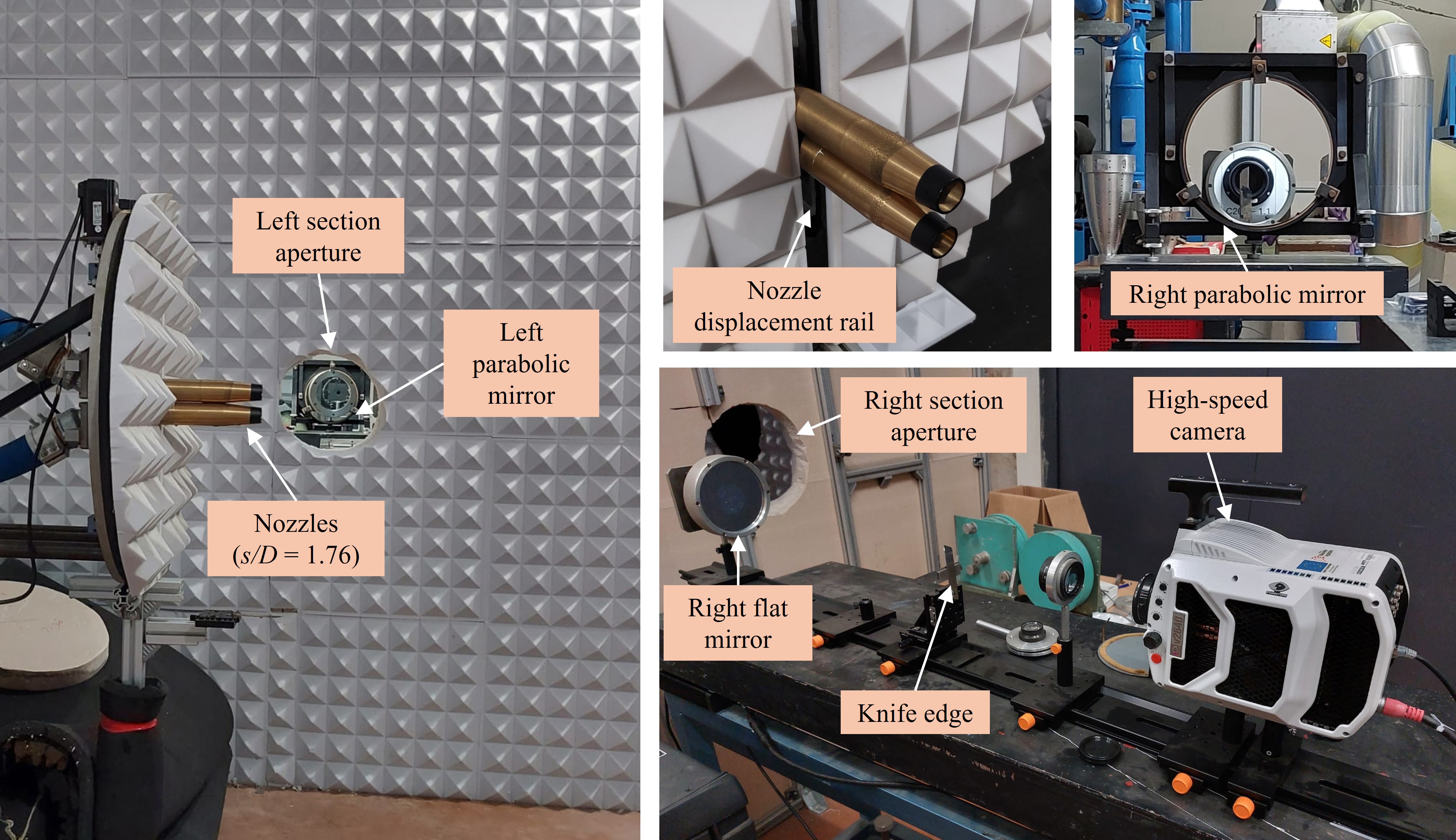}}
\caption{Experimental setup and elements of the high-speed schlieren measurement system.}
\label{fig:exp_setup}
\end{figure}

High-speed schlieren visualizations and low-speed PIV measurements of the twin-jet flow field have been performed at the T200 wind tunnel facility of the PROMETEE platform of Institut Pprime (CNRS - Universit\'{e} de Poitiers - ISAE-ENSMA), France. The setup is shown in figure~\ref{fig:exp_setup}. The flow is generated using a $200$ bar compressed air network that can reach operational conditions up to an isentropic Mach number $M_j = 2$ for the twin-jet configuration. A heating system based on a series of tanks with heated nickel balls is used to increase and maintain the total temperature of the air reaching the nozzles. The left image of figure~\ref{fig:exp_setup} shows the twin-jet system situated in an anechoic chamber.

Schlieren visualizations are obtained using a classical Z-type setup, for which some components are also illustrated in figure~\ref{fig:exp_setup}. A continuous light source is generated by a 60 W LED, which passes through an aperture that prevents direct light from the source to enter the test section. Two parabolic mirrors, each 30 cm in diameter and with a 3 m focal length, produce a collimated light beam that travels through the test section in the $z$ direction, according to the reference frame in figure~\ref{fig:twinjet_geometry}. To form a Z-shaped optical path that allows for a more compact experimental setup, two additional flat mirrors, each 12 cm in diameter, are incorporated. A vertical knife edge is positioned at the focal length of the second parabolic mirror, making the resulting light intensity field recovered in the schlieren images proportional to the streamwise density gradients in the flow.

\begin{figure}
\centerline{\includegraphics[width=0.99\textwidth]{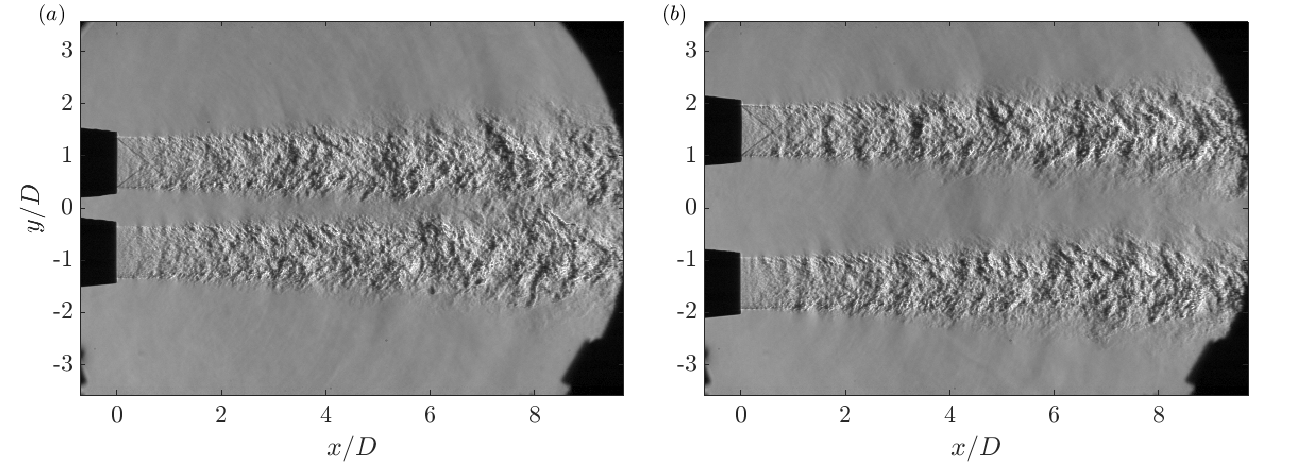}}
\caption{Instantaneous schlieren snapshots of the twin-jet flow field at $M_j = 1.54$: $(a)$ $s/D = 1.76$; $(b)$ $s/D = 3$. The dark regions near the downstream boundary of the images correspond to the edges of the right parabolic mirror.}
\label{fig:schlieren_snapshots}
\end{figure}

A Phantom v2640 high-speed camera was used to record the images. For the current datasets, $N_s = 30000$ instantaneous snapshots were recorded at a sampling frequency of $f_s = 68$ kHz ($\Delta t = 1.47 \times 10^{-5}$ s) and a spatial resolution of $352 \times 512$ pixels. The Strouhal number corresponding to the recording Nyquist frequency is $St_\textrm{Ny} = f_s D / (2 u_j) = 1.93$ ($2 \Delta t u_j / D = 0.52$), which is high enough to resolve the coherent structures associated with mixing noise, which develop in the range $St \approx [0.1, 1]$ for supersonic jets~\citep{Sinha:JFM14}. The total number of convective time units spanned by the recording is $\tau c_\textrm{KH} / D \approx 5440$, where $\tau = \Delta t N_s = 0.44$ s denotes the total recording time and $c_\textrm{KH} = 0.7u_j$ is the estimated convective velocity of Kelvin-Helmholtz structures. Figure~\ref{fig:schlieren_snapshots} illustrates two instantaneous schlieren snapshots respectively recorded for $s/D = 1.76$ and $s/D = 3$.

PIV measurements are performed to obtain a mean velocity field to validate the RANS calculations. The PIV measurement system consists of two side-by-side LaVision Imager LX 16M cameras which record 500 instantaneous, high-resolution images (dual frames) at a sampling frequency of 4.4 Hz and at a resolution of 16 megapixels each, employing a time delay between frames of 1.5 $\upmu s$. The flow is seeded with pressurized glycerin vapour in the settling chamber, prior to nozzle expansion, while the ambient region is seeded with a fog generator employing a glycerin-water mixture. A double-pulse Nd-YAG laser operating at 532 nm is used, which illuminates the seeded particles in the $z = 0$ plane, enabling measurement of the mean-flow velocity in the symmetry plane containing the two jets. The PIV measurement window covers the region $(x/D,y/D) \in [0,10]\times[-5,5]$, extending downstream approximately up to the end of the potential core. After processing, the resulting instantaneous velocity fields have a resolution of $695 \times 622$ pixels.

\subsection{Eduction of coherent structures from schlieren visualizations}
\label{sec:SPOD_method}

In order to validate the wavepackets modelled by plane-marching PSE, coherent structures are extracted by means of spectral proper orthogonal decomposition (SPOD)~\citep{Lumley1970,Towne2018} applied to the high-speed schlieren visualizations. The coherent structures educed by SPOD in subsonic and perfectly-expanded supersonic turbulent jets are known to represent wavepackets related to the Kelvin-Helmholtz instability of the shear layers and non-modal dynamics including Orr and lift-up mechanisms~\citep{Suzuki:JFM06,Gudmundsson:JFM11,Tissot:PRF2017,Jordan:AIAAC2017,Sasaki:JFM17,Schmidt:JFM2018,Lesshafft:PRF2019,Nogueira:JFM2019,Pickering:JFM2020}.

The application of SPOD to schlieren visualizations of twin jets has proved successful in obtaining empirical descriptions of the screech phenomenon~\citep{Edgington-Mitchell2022}, which involves a highly organized structure at localized tonal frequencies. In the broadband, amplifier system constituted by the perfectly-expanded twin-jet flow field, small-scale vortical structures are emphasized by the schlieren visualizations, making the extraction of coherent structures difficult: a large number of snapshots is necessary to reach converged SPOD modes. For this reason, instead of directly using the schlieren images to educe the coherent structures, a different quantity derived from the schlieren field is used to build the cross-spectral density (CSD) matrix.

Doak's momentum potential theory~\citep{Doak1989,Jordan:JSV2013} can be employed to extract the irrotational momentum potential component associated with the line-of-sight integrated momentum density field~\citep{Prasad2022}, denoted by $\Theta$, from the schlieren field, denoted by $\sigma$. As illustrated by~\citet{PadillaMontero2024}, for a given dataset, using $\Theta$ instead of $\sigma$ to build the CSD matrix in SPOD allows for a more effective extraction of coherent structures, facilitating comparison with wavepacket models based on linear stability theory. The irrotational field acts as a filter that removes much of the small-scale turbulence fluctuations from the schlieren images, reducing the dimensional complexity of the dataset.

The same methodology described in~\citet{PadillaMontero2024} is here employed for educing coherent structures. Doak's momentum potential theory yields a relationship between the density fluctuation field ($\rho'$) and the momentum potential fluctuation field ($\psi'$) in the form of a linear Poisson equation~\citep{Doak1989}:

\begin{equation}
\frac{\partial^2 \psi'}{\partial x_i^2} = \frac{\partial \rho'}{\partial t}.
\label{eq:Poisson}
\end{equation}

\noindent Following~\citet{Prasad2022}, equation~\eqref{eq:Poisson} can be differentiated with respect to $x$, yielding:

\begin{equation}
\frac{\partial^2}{\partial x_i^2} \left( \frac{\partial \psi'}{\partial x} \right) = \frac{\partial }{\partial t} \left( \frac{\partial \rho '}{\partial x} \right),
\label{eq:Poisson_dx}
\end{equation}

\noindent and can be integrated along $z$ (the line of sight of the schlieren measurements) to obtain:

\begin{equation}
\frac{\partial^2 \Theta'}{\partial x_i^2} = \frac{\partial \sigma'}{\partial t},
\label{eq:Poisson_sch}
\end{equation}

\noindent where $\sigma' = \int (\partial \rho' / \partial x)\,\mathrm{d}z$ denotes the schlieren fluctuation field and $\Theta' = \int (\partial \psi' / \partial x)\,\mathrm{d}z$ is the line-of-sight integrated streamwise derivative of the momentum potential fluctuation. Equation~\eqref{eq:Poisson_sch} can be solved using \textit{ad hoc} boundary conditions on the truncated domain given by the schlieren images, allowing calculation of the $\Theta'$ field associated with the experimental measurements. Here, solution of the Poisson equation is incorporated into the SPOD algorithm, and $\Theta'$ is computed in the spectral domain. Special care is taken to filter out spurious harmonic components that arise from solution of the Poisson equation in a truncated domain with unknown boundary conditions. The reader is referred to~\citet{PadillaMontero2024} for details of the solution approach.

To facilitate comparison of PM-PSE modes against experimentally-educed structures, the symmetry of the system along the $y = 0$ plane is also exploited in the treatment of the experimental datasets. Before solving the SPOD eigenvalue problem, two new sets of $\Theta'$ realizations are computed by splitting each $\hat{\Theta}$ snapshot by the line at $y = 0$, namely, a symmetric set $\hat{\Theta}_s = (\hat{\Theta}_u + \hat{\Theta}_l)/2$ and an antisymmetric one $\hat{\Theta}_a = (\hat{\Theta}_u - \hat{\Theta}_l)/2$, with $\hat{\Theta}_u$ and $\hat{\Theta}_l$ respectively denoting the upper ($y > 0$) and lower ($y < 0$) halves of the original realizations. The SPOD eigenvalue problem is then solved for the symmetric and antisymmetric datasets separately. This decomposition is analogous to the $D_2$ decomposition~\citep{Sirovich:PoF1990} employed for the study of rectangular twin jets~\citep{Yeung:AIAAC2022}, which allows symmetric and antisymmetric fluctuations to be extracted without loss of generality. This permits a systematic comparison with the wavepackets computed using PM-PSE featuring symmetric/antisymmetric boundary conditions.

Each schlieren dataset is divided into 57 blocks of 1024 snapshots, with an overlap of 50\%, yielding frequency bins of $\Delta f = 66$ Hz ($\Delta St = 0.0038$). A temporal Hamming window is used to reduce spectral leakage. The SPOD decomposition of the momentum potential fluctuation field yields SPOD modes in terms of $\Theta'$. To compare the educed coherent structures with the PM-PSE wavepackets, SPOD modes expressed in terms of schlieren fluctuations are also of interest. These SPOD fields can be computed as an extended SPOD problem (see for example~\citet{FreundColonius:AIAA02,Boree2003,Sinha:JFM14,Souza:JFM19,Himeno:JFM21,Karban2023}). In the results shown later in this work, two different flow-field variables are shown for each SPOD mode, namely, the $\hat{\Theta}$ field and the $\hat{\sigma}_\Theta$ field, which denotes the schlieren fluctuation field corresponding to the coherent structure educed based on the cross-spectral density of $\Theta'$.

\section{Results}
\label{sec:results}

\subsection{Mean flow}

\begin{figure}
\centerline{\includegraphics[width=0.99\textwidth]{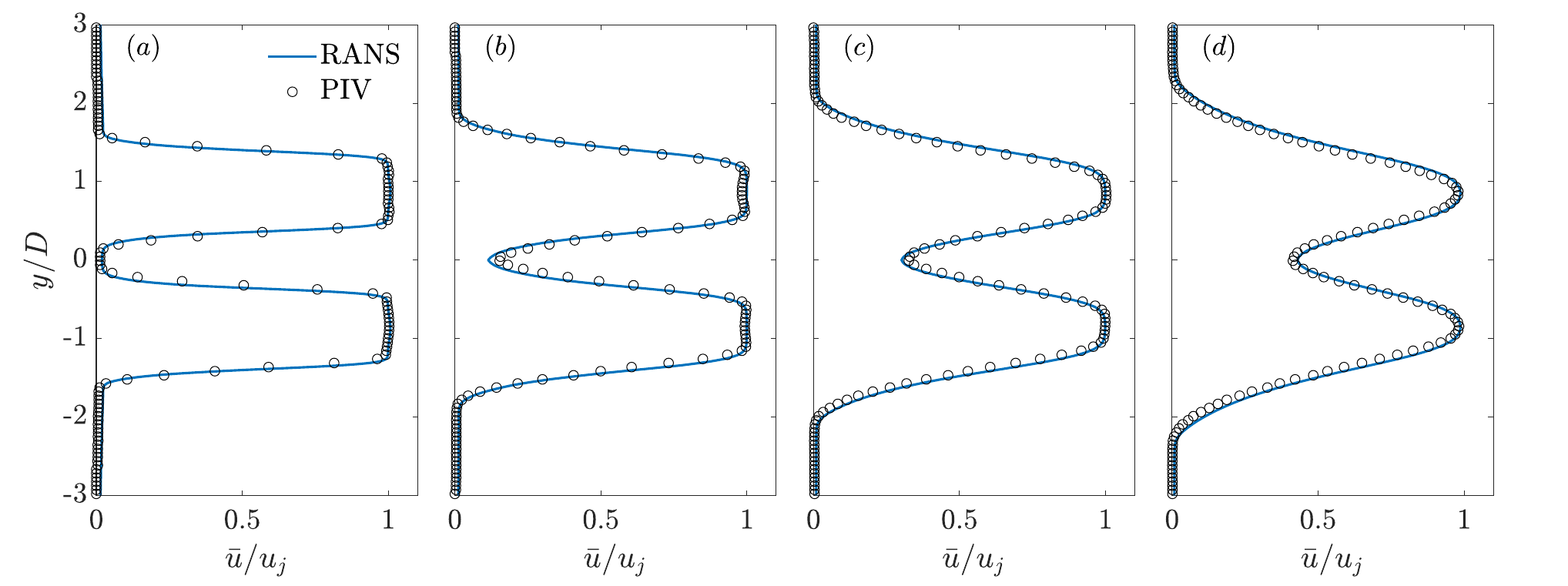}}
\caption{Comparison of mean streamwise velocity profiles along $y$ (and at $z = 0$) for four different streamwise locations, between the RANS solution and the PIV mean flow for $s/D = 1.76$: $(a)$ $x/D = 2$; $(b)$ $x/D = 4$; $(c)$ $x/D = 6$; $(d)$ $x/D = 8$.}
\label{fig:RANS_vs_PIV_vs_x}
\end{figure}

\begin{figure}
\centerline{\includegraphics[width=0.99\textwidth]{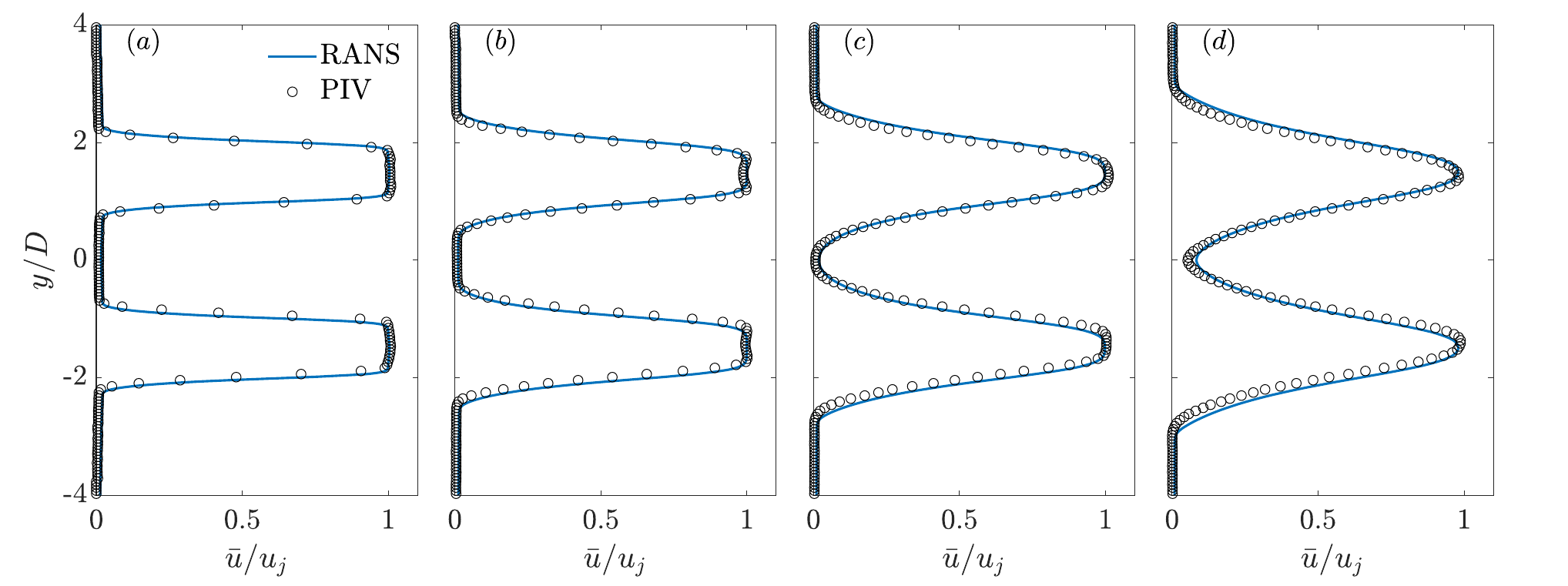}}
\caption{Comparison of mean streamwise velocity profiles along $y$ (and at $z = 0$) for four different streamwise locations, between the RANS solution and the PIV mean flow for $s/D = 3$: $(a)$ $x/D = 2$; $(b)$ $x/D = 4$; $(c)$ $x/D = 6$; $(d)$ $x/D = 8$.}
\label{fig:RANS_vs_PIV_vs_x_sD3}
\end{figure}

First, the RANS mean flow employed for the PSE calculations is discussed and compared with the PIV measurements. Figures ~\ref{fig:RANS_vs_PIV_vs_x} and ~\ref{fig:RANS_vs_PIV_vs_x_sD3} show the mean streamwise velocity profiles obtained from the RANS computations at the symmetry plane $z = 0$. 
For each jet separation, the RANS and PIV fields are in good agreement. The RANS model accurately captures the mixing-layer thickness, the centerline velocity and the velocity increase in the region between the two jets. The profiles highlight the interaction between the jets that takes place at the merging region between the inner shear layers. In this region, the flow velocity increases above that of the equivalent isolated jet, making the velocity gradient (shear magnitude) in the inner mixing layers smaller with respect to that of the external one. This effect reflects the loss of axisymmetry, which is significantly stronger for $s/D = 1.76$. Small discrepancies are progressively found for $x/D > 6$, which manifest as a small overprediction of the external mixing layer thickness and of the velocity magnitude in the region between the jets.

\begin{figure}
\centerline{\includegraphics[width=0.99\textwidth]{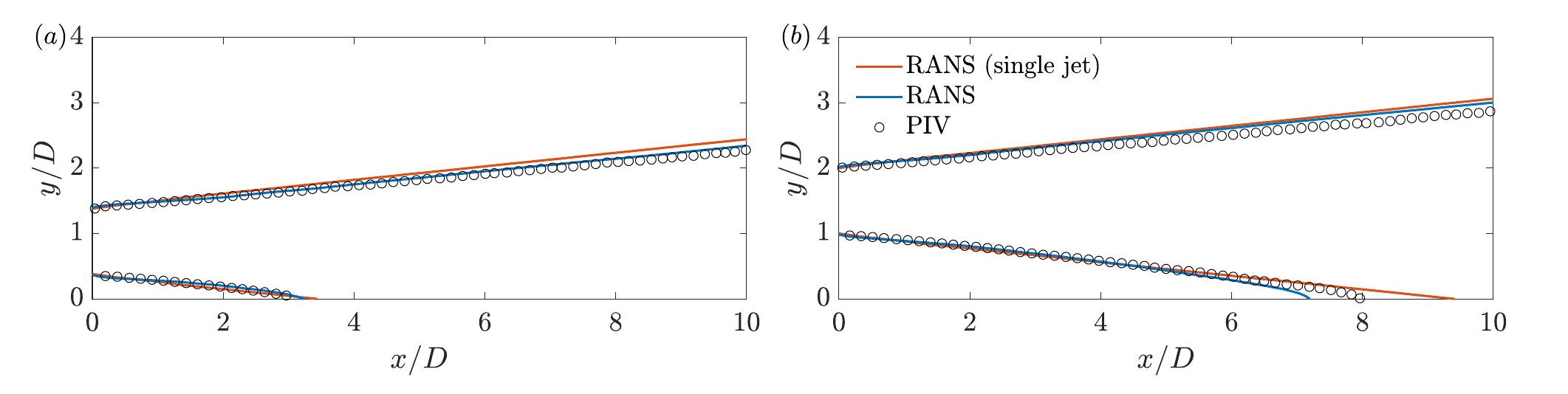}}
\caption{Comparison of mean streamwise velocity contours corresponding to the external shear-layer boundary ($\bar{u}/u_j = 0.05$) at $z = 0$ between RANS and PIV: $(a)$ $s/D = 1.76$; $(b)$ $s/D = 3$. The single jet (axisymmetric) RANS solution is also added for comparison.}
\label{fig:RANS_vs_PIV_shear_layer}
\end{figure}

Figure~\ref{fig:RANS_vs_PIV_shear_layer} shows the evolution of the inner and outer shear-layer boundaries of the twin jet for each spacing in both the RANS and PIV fields. This is defined by the $u/u_j = 0.05$ velocity contour. As discussed above, the comparison against the PIV measurements reveals a small overprediction of the RANS mixing layer thickness which increases further downstream. The axisymmetric (single jet) solution is also shown for comparison. In the twin-jet system, the spread rate of the outer mixing layer is reduced with respect to that of the isolated jet, while the spreading of the inner mixing layer is increased owing to entrainment of quiescent fluid in the region between the jets. These trends are consistent with the findings of previous work~\citep{Moustafa:AIAAJ1994,Alkislar:AIAAJ2005,Goparaju2018}.

\begin{figure}
\centerline{\includegraphics[width=0.99\textwidth]{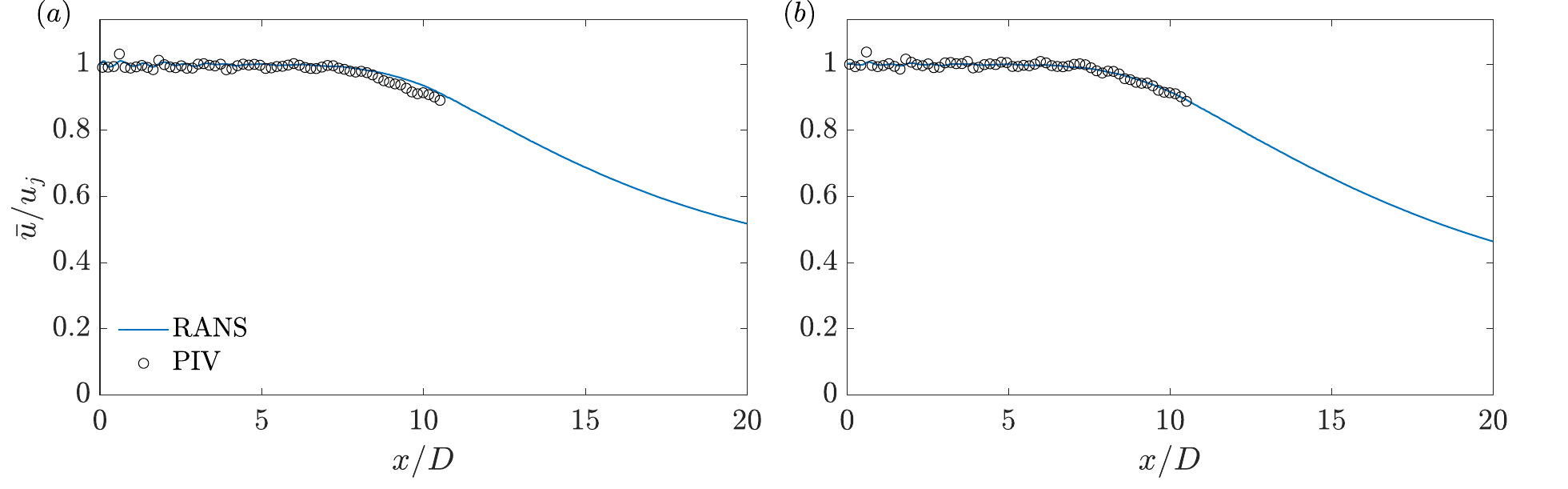}}
\caption{Comparison of mean streamwise velocity profiles along the top nozzle axis between the RANS solution and the PIV mean flow for $(a)$ $s/D = 1.76$ and $(b)$ $s/D = 3$.}
\label{fig:RANS_vs_PIV_nzz_axis}
\end{figure}

Figure~\ref{fig:RANS_vs_PIV_nzz_axis} shows the streamwise velocity evolution along the top nozzle axis for each $s/D$, illustrating the decay of the centerline velocity near and after the end of the potential core. The effect is intimately related to the turbulent diffusion, and constitutes an additional validation of the numerical solution.

The mean-flow results demonstrate the ability of the 3D RANS solution to accurately model the twin-jet mean flow, effectively accounting for the non-linear mean-flow interaction between the jets. This overcomes the shortcomings of more simplified models employed in previous work, such as the tailored twin-jet mean flow constructed by the linear superposition of two isolated jets~\citep{Rodriguez2018,Rodriguez2021,Rodriguez2023JFM}.

\subsection{Twin-jet wavepackets modelled with PM-PSE}
\label{sec:PSE_wavepackets}

PM-PSE calculations have been performed in the frequency range $St = [0.1,1]$. For each frequency, four different disturbances have been marched downstream, denoted by SS0, SA0, SS1 and SA1~\citep{Rodriguez2023JFM}. The first letter refers to the symmetry with respect to $z = 0$, the second letter to the symmetry with respect to $y = 0$ (as described in section~\ref{sec:PSE_numerics}), and the number denotes whether the disturbance is of toroidal nature (equivalent to $m = 0$ in a single round jet) or of flapping nature (equivalent to $m = 1$ in a single round jet). Disturbances antisymmetric with respect to the symmetry plane at $z = 0$ are not treated in this work as they cannot be characterized by the schlieren visualizations. Higher modes (twin-jet analogous for $m > 1$), although being recovered by the two-dimensional linear stability theory, are not considered as they are more challenging to educe from the experimental visualizations.

\begin{figure}
\centerline{\includegraphics[width=0.99\textwidth]{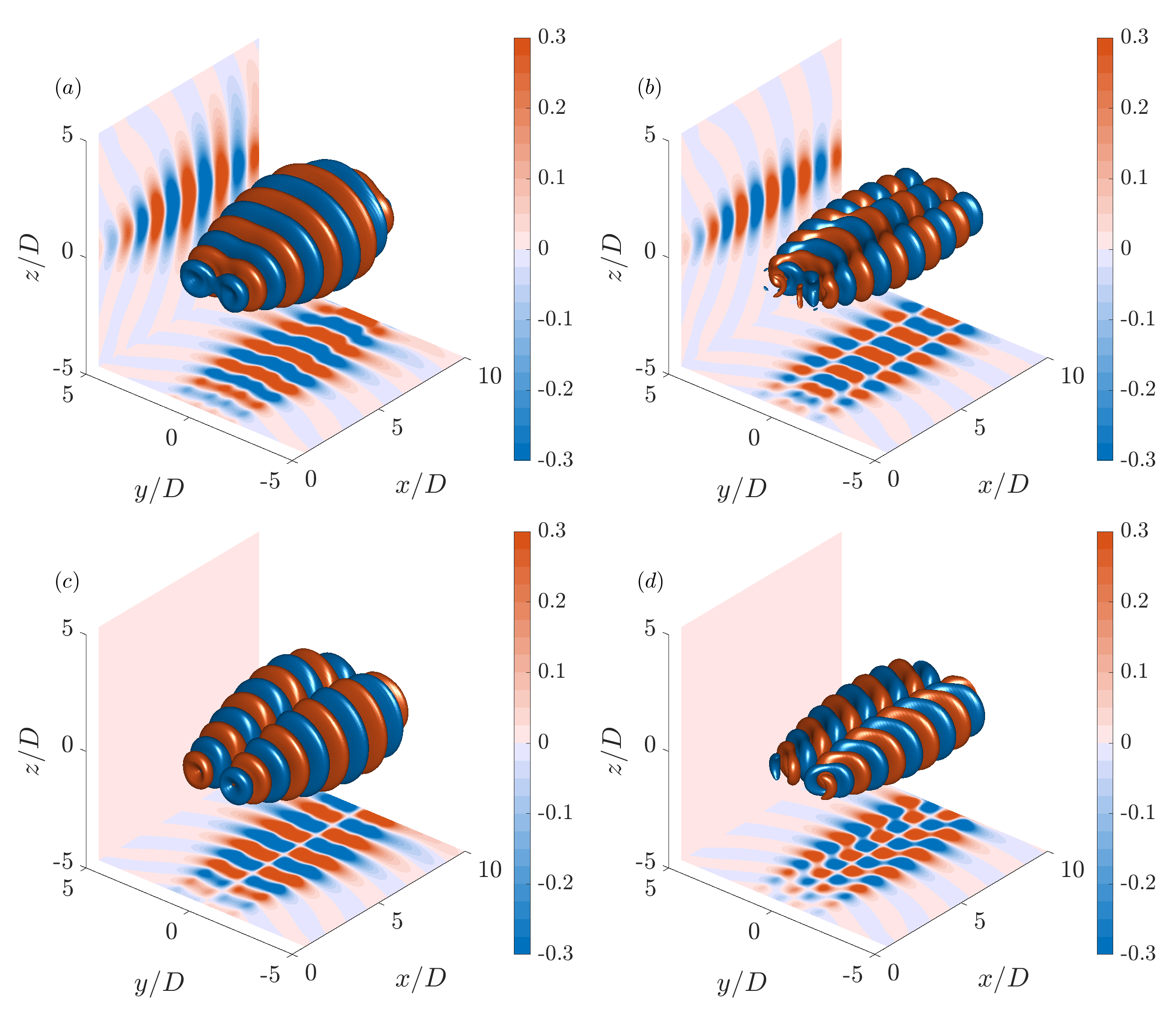}}
\caption{Isosurfaces of the real part of the pressure fluctuation for PM-PSE modes $(a)$ SS0, $(b)$ SS1, $(c)$ SA0 and $(d)$ SA1 at $St = 0.4$, $s/D = 1.76$. Values are normalized with respect to the maximum absolute value of the real part of $\hat{p}$. Displayed isosurfaces correspond to $\Real \{ \hat{p} \} = 0.1$ (orange) and $\Real \{ \hat{p} \} = -0.1$ (blue). The projected filled contours correspond to the real part of the pressure fluctuation in the $xy$ symmetry plane located at $z = 0$ and the $xz$ symmetry plane located at $y = 0$. The colorbars refer to the projected contours.}
\label{fig:surf3D_PMPSE_all_St0d4}
\end{figure}

\begin{figure}
\centerline{\includegraphics[width=0.99\textwidth]{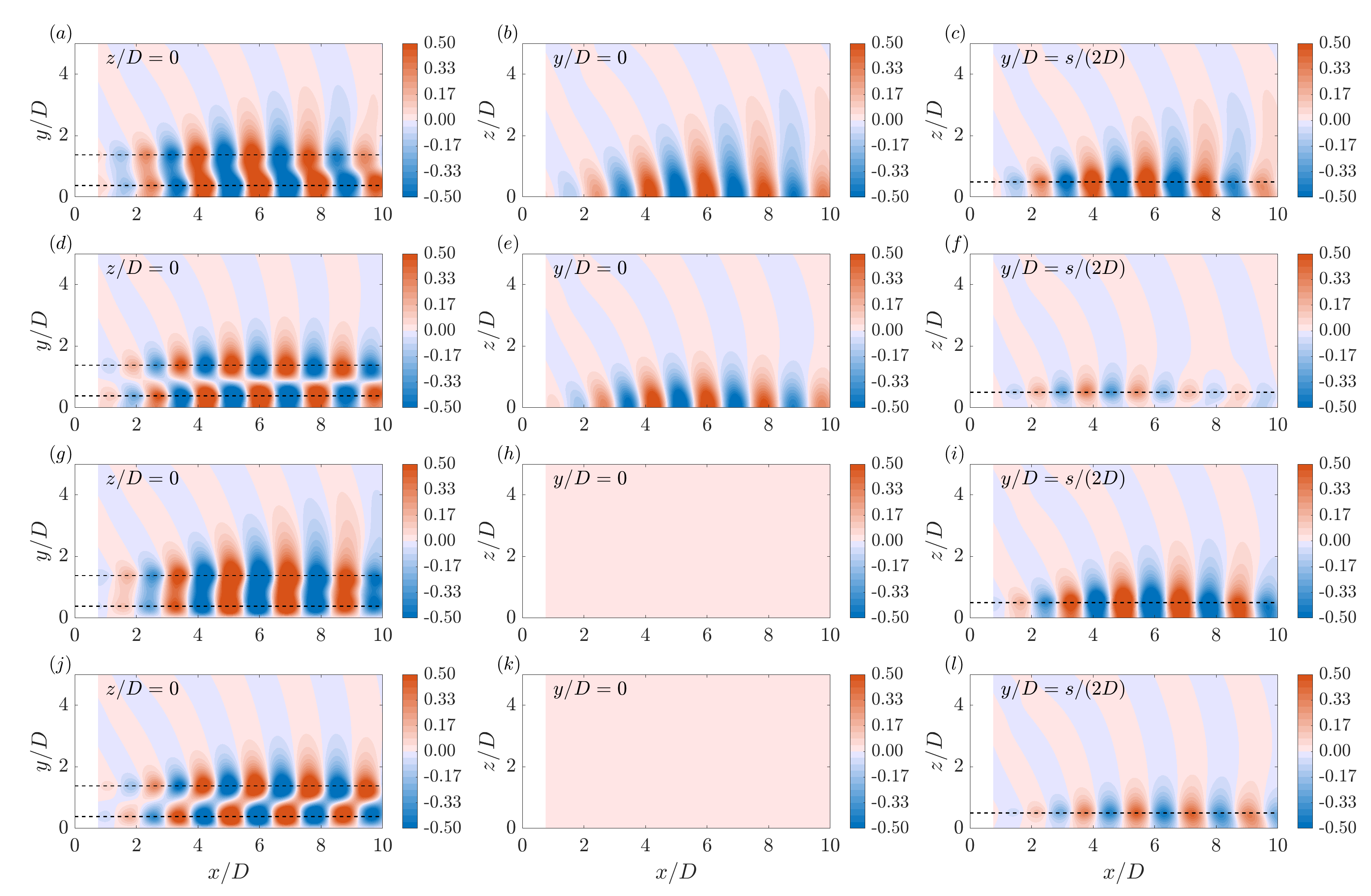}}
\caption{Contours of the real part of the pressure fluctuation for the different PM-PSE modes at $St = 0.4$, $s/D = 1.76$: $(a,d,g,j)$ symmetry plane at $z/D = 0$; $(b,e,h,k)$ symmetry plane at $y/D = 0$; $(c,f,i,l)$ nozzle mid-plane $y/D = s/(2D)$; $(a,b,c)$ mode SS0; $(d,e,f)$ mode SS1; $(g,h,i)$ mode SA0; $(j,k,l)$ mode SA1. Values are normalized with respect to the maximum absolute value of the real part of $\hat{p}$ in the entire field. Only one quarter of the twin-jet system is displayed according to the two inherent symmetry planes. The black dashed lines denote the nozzle lip lines.}
\label{fig:cutplanes_PMPSE_all_St0d4}
\end{figure}

\begin{figure}
\centerline{\includegraphics[width=0.99\textwidth]{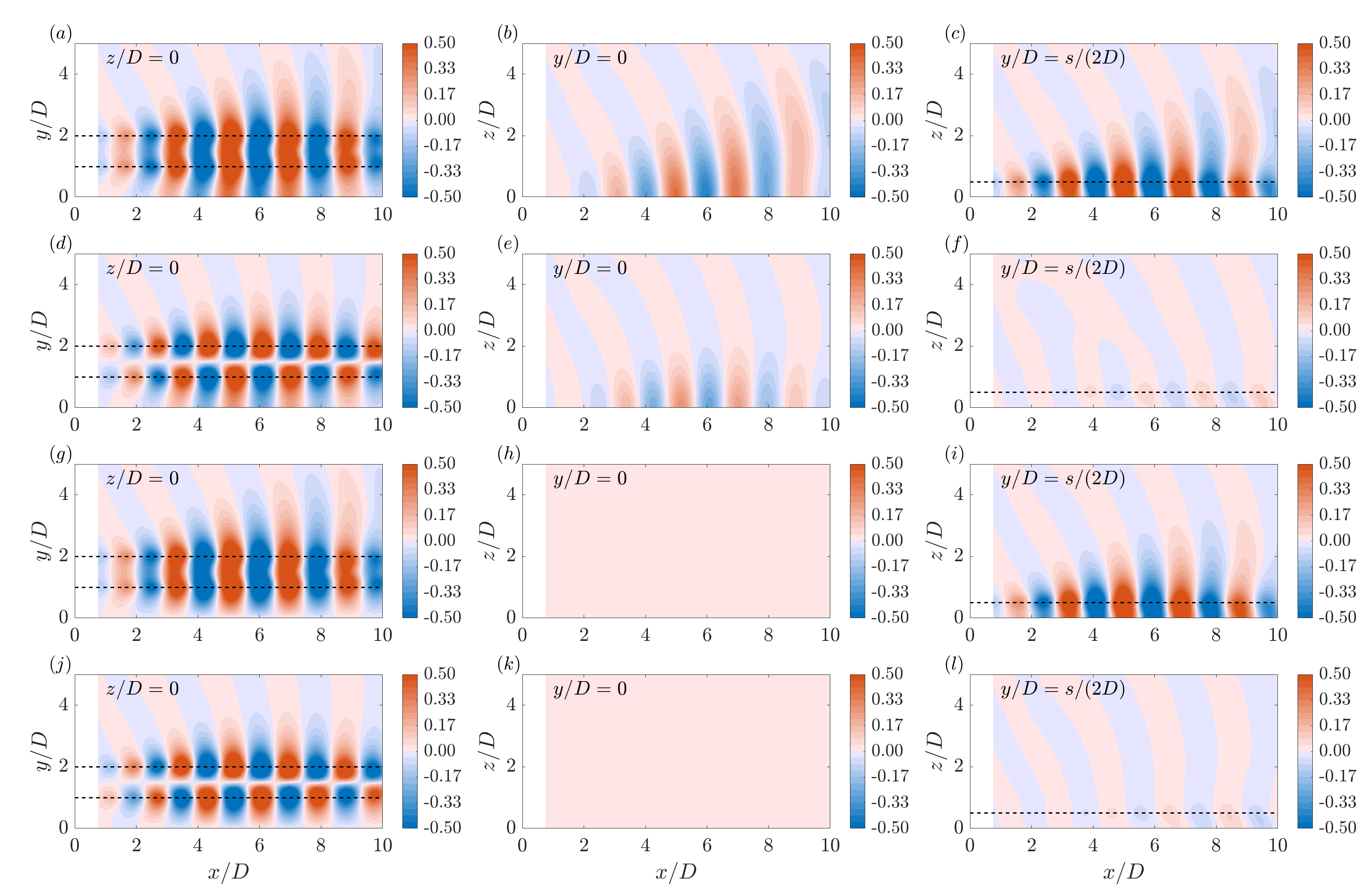}}
\caption{Contours of the real part of the pressure fluctuation for the different PM-PSE modes at $St = 0.4$, $s/D = 3$: $(a,d,g,j)$ symmetry plane at $z/D = 0$; $(b,e,h,k)$ symmetry plane at $y/D = 0$; $(c,f,i,l)$ nozzle mid-plane $y/D = s/(2D)$; $(a,b,c)$ mode SS0; $(d,e,f)$ mode SS1; $(g,h,i)$ mode SA0; $(j,k,l)$ mode SA1. Values are normalized with respect to the maximum absolute value of the real part of $\hat{p}$ in the entire field.}
\label{fig:cutplanes_PMPSE_all_St0d4_s3}
\end{figure}

Figure~\ref{fig:surf3D_PMPSE_all_St0d4} shows the three-dimensional structure of the four PM-PSE wavepackets computed for $St = 0.4$ and $s/D = 1.76$ (closely-spaced jets). For each mode, isosurfaces of the real part of the pressure fluctuation are shown, together with projected filled contours that correspond to the two symmetry planes at ($y = 0$) and ($z = 0$). The isosurfaces clearly illustrate the toroidal nature of modes SS0 and SA0, characterized by ring-like structures that grow downstream within the potential core and the mixing layers of each jet, while the projected contours serve to highlight the significant Mach-wave radiation associated with the wavepacket structure. The three-dimensional structure of modes SS1 and SA1 reveals flapping fluctuations featuring a change of sign across the axis of each jet, and which also radiate to the far-field in both $y$ and $z$ directions.

To better highlight some key features of the three-dimensional structure of the twin-jet wavepackets, figures~\ref{fig:cutplanes_PMPSE_all_St0d4} and \ref{fig:cutplanes_PMPSE_all_St0d4_s3} present filled contours of the real part of the pressure fluctuation along three different cutplanes for $s/D = 1.76$ and $s/D = 3$, respectively. In addition to both twin-jet symmetry planes, the $xz$ plane located at $y = s/(2D)$ (nozzle axis) is also shown for comparison. For both nozzle spacings, mode SS0 features similar amplitude levels and Mach-wave radiation signatures in both the $z = 0$ and $y = s/(2D)$ planes. In contrast, for mode SA0, while the wavepacket radiates in the $z = 0$ plane up to the end of the displayed streamwise range ($x/D = 10$), the Mach waves in the $y = s/(2D)$ plane are found to be weaker and the radiation begins to decrease from $x/D = 8$ onwards, indicating a predominance of sideline radiation for this mode. Similarly, both flapping modes also exhibit stronger radiation signatures in the $z/D = 0$ plane than in the $y/D = s/(2D)$ plane, which in this case is expected owing to the near-zero amplitude found for these fluctuations along the axis of each nozzle.

For closely-spaced jets ($s/D = 1.76$), significant differences are found in the streamwise evolution of the wavepackets between the inner and the outer mixing layers of each jet. This is clearly visible in the $z/D = 0$ contour plots (figure~\ref{fig:cutplanes_PMPSE_all_St0d4}$(a,d,g,j)$). For the toroidal fluctuations, this makes the ring-like structures appear tilted with respect to the axis of the jet, rather than perpendicular to it as would be expected in the axisymmetric case. For the flapping modes, it causes the fluctuations to not be in perfect counter-phase across the nozzle axis, preventing the amplitude at the jet axis from reaching a zero value. As shown below, these phase differences are associated with a mismatch in the phase-speed evolution $c_{ph}$ of the Kelvin-Helmholtz waves along the inner and outer mixing layers, which is attributed to the mean-flow interaction taking place in the region between the two jets.

\begin{figure}
\centerline{\includegraphics[width=0.99\textwidth]{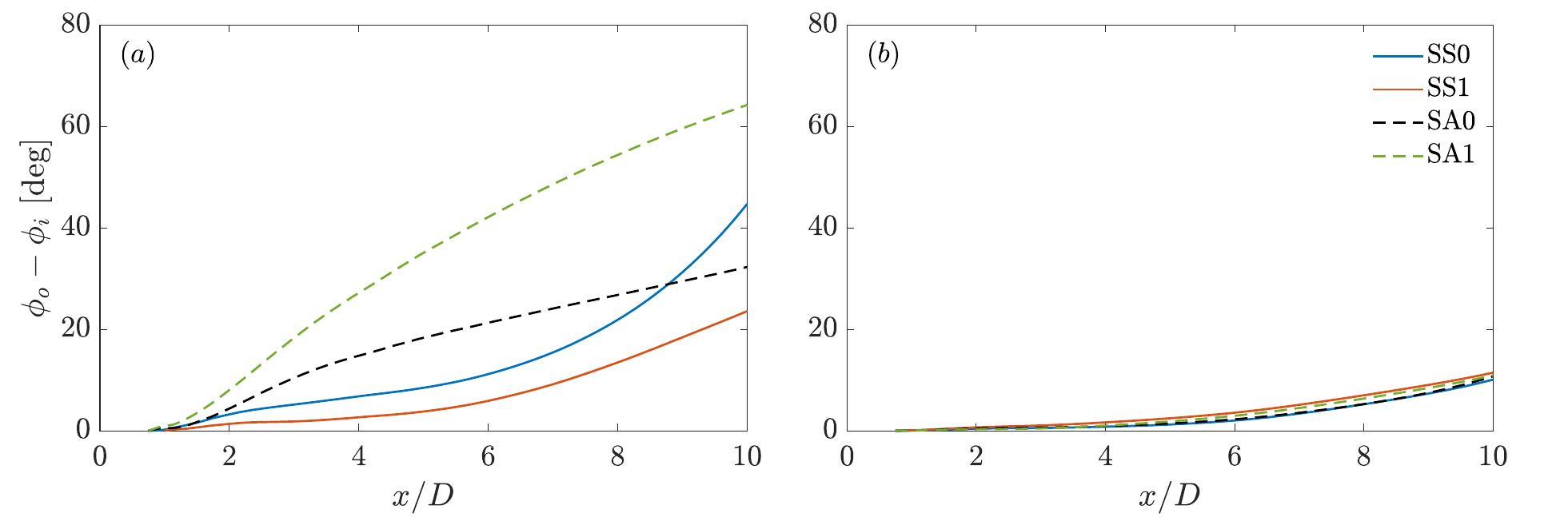}}
\caption{Streamwise evolution of the phase difference between the outer and inner lip lines of the pressure fluctuation for the different PM-PSE modes ($St = 0.4$): $(a)$ $s/D = 1.76$; $(b)$ $s/D = 3$. The phase along each line is referenced to the initial streamwise station where the PSE marching is initialized ($\phi_0$). $\phi_o$ and $\phi_i$ respectively denote the phase along the outer and inner lip lines.}
\label{fig:phase_diff_vs_x_St0d4}
\end{figure}

Figure~\ref{fig:phase_diff_vs_x_St0d4} shows the phase difference in the pressure fluctuation of the four PM-PSE modes between the inner and outer lip lines of each nozzle (represented by the black dashed lines in the $z/D=0$ plane in figures~\ref{fig:cutplanes_PMPSE_all_St0d4} and \ref{fig:cutplanes_PMPSE_all_St0d4_s3}). The phase extracted along each lip line is represented relative to the phase at the initial position of the PM-PSE marching ($\phi_0$ at $x/D = 0.75$). For $s/D = 1.76$, all four modes accumulate a notable phase difference near the end of the potential core ($x/D \approx 10$), with mode SA1 exhibiting the largest phase shift between the inner and outer lip lines, which exceeds 60 degrees.

\begin{figure}
\centerline{\includegraphics[width=0.99\textwidth]{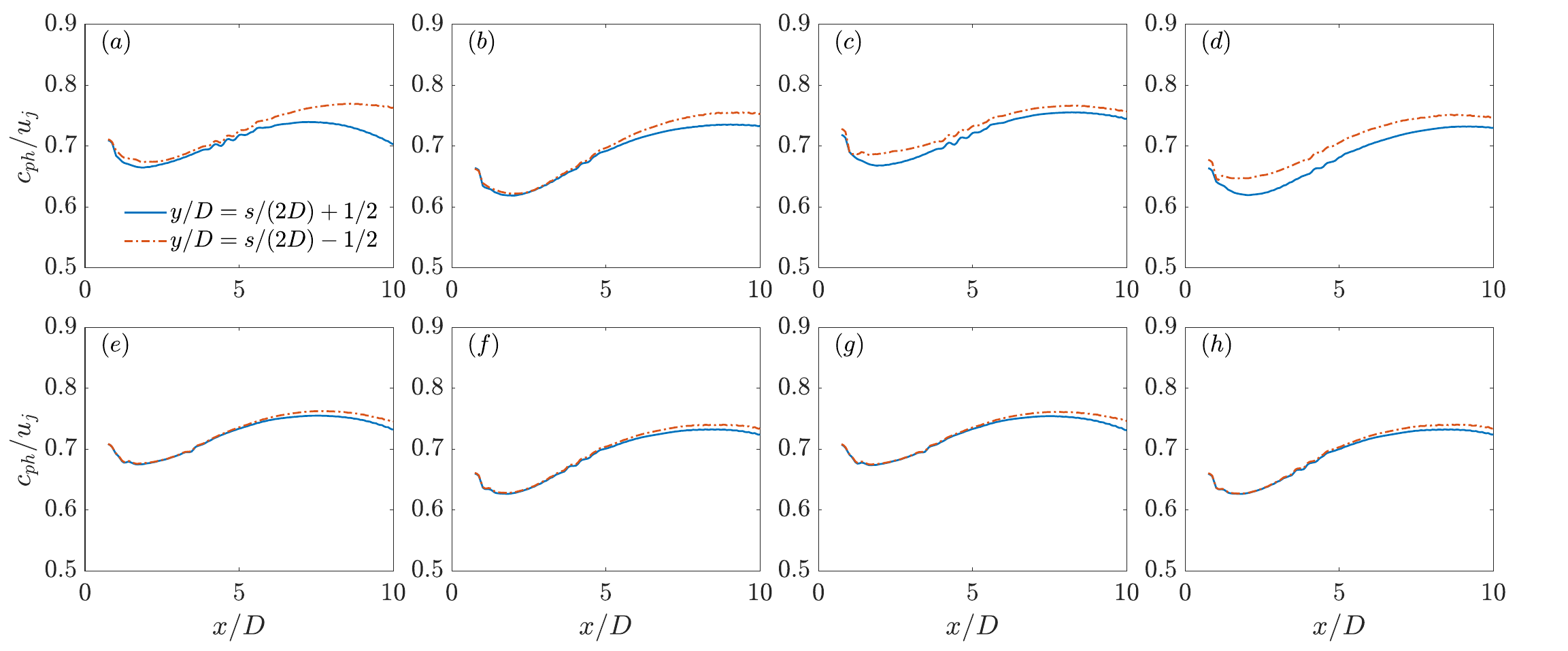}}
\caption{Streamwise evolution of the phase speed of the pressure fluctuation along the outer and inner lip lines for the different PM-PSE modes at $St = 0.4$: $(a$-$d)$ $s/D = 1.76$; $(e$-$h)$ $s/D = 3$; $(a,e)$ SS0; $(b,f)$ SS1; $(c,g)$ SA0; $(d,h)$ SA1.}
\label{fig:cph_vs_x_St0d4}
\end{figure}

Figure~\ref{fig:cph_vs_x_St0d4} presents the streamwise evolution of the phase speed of the pressure fluctuation for each mode along the lip lines. To calculate $c_{ph}$, the streamwise wavenumber of the fluctuation $k_x$ at a given $y$ location is obtained by computing the streamwise derivative of the phase with respect to $x$. The comparison between the phase-speed evolution along each line reveals a higher phase speed in the inner lip line with respect to the outer one. For modes SS0 and SS1, the phase-speed evolution correlates with the streamwise mean-flow velocity evolution at these positions. In particular, SS0 and SS1 exhibit an important phase-speed mismatch for $x/D > 5$, which is the streamwise range in which the mean-flow velocity along the inner lip line becomes significantly higher than along the outer lip line. Therefore, the higher streamwise velocity found in the inner mixing layers leads to a larger convective speed for the wavepackets in this region. On the other hand, the antisymmetric modes SA0 and SA1 present a significant discrepancy in $c_{ph}$ between both lines for the entire streamwise range. This is attributed to the combination of two effects: the higher mean-flow velocity in the inner mixing layer, as for the symmetric modes, but also the fact that the antisymmetric wavepacket structures contained in the inner mixing layer are confined between the nozzle axis and the symmetry plane. For low $St$, the spatial wavelength of the fluctuations becomes comparable to the spacing between the nozzle axis and the symmetry axis. This has an impact in the development of the antisymmetric structures, modulating the wavenumber in the inner mixing layer such that the phase speed is increased in this case.

The phase-speed mismatch between the inner and outer mixing layers is found to be much weaker for higher frequencies ($St > 0.4$), as the radial reach of the wavepacket structures across the plane between jets decreases with increasing frequency. For lower frequencies ($St \leq 0.4$), the antisymmetric flapping modes (SA1) remain the fluctuations which manifest the largest distortion owing to the phase-speed difference, in line with the results presented in figure~\ref{fig:phase_diff_vs_x_St0d4}$(a)$. More details on the evolution of SA1 modes with $St$ are provided in section \ref{sec:inconsistent_chi}.

In the case of larger jet spacing ($s/D = 3$), the phase-speed mismatch and associated phase difference between the inner and outer mixing layers is much weaker, as shown in figures~\ref{fig:phase_diff_vs_x_St0d4}$(b)$ and~\ref{fig:cph_vs_x_St0d4}$(e$-$h)$. In this configuration, the topology of the modes is closer to the axisymmetric, single-jet behaviour, and the structure of the wavepackets features a perpendicular alignment with respect to the nozzle axis (see figure~\ref{fig:cutplanes_PMPSE_all_St0d4_s3}).

\subsection{Coherent structures educed from the schlieren measurements}
\label{sec:SPOD_results}

\begin{figure}
\centerline{\includegraphics[width=0.99\textwidth]{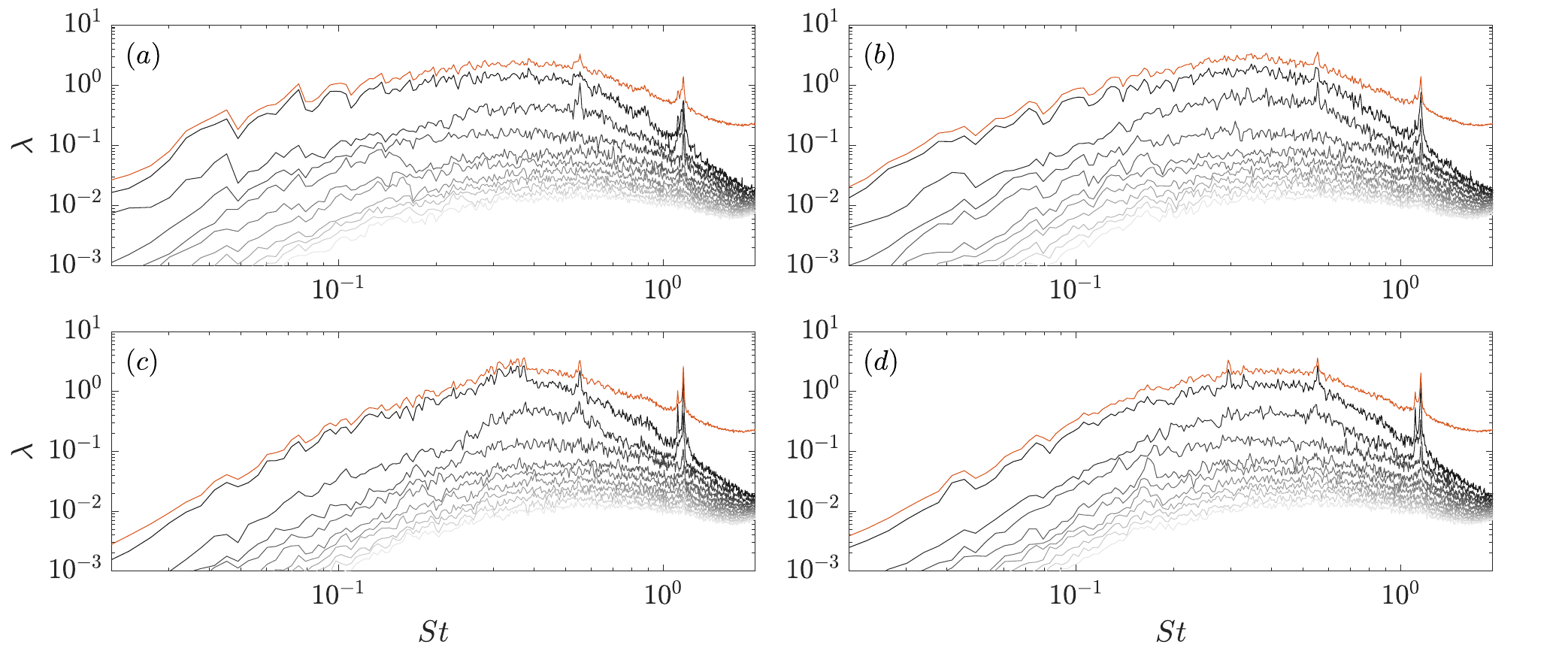}}
\caption{SPOD spectra obtained using the cross-spectral density of $\Theta$ fluctuations: $(a,c)$ $s/D = 1.76$; $(b,d)$ $s/D = 3$; $(a,b)$ symmetric datasets; $(c,d)$ antisymmetric datasets. Only the first ten SPOD modes are shown, ranked following a gray scale between mode 1 (black) and mode 10 (white). $\lambda$ represents the spectral density (eigenvalue of the cross-spectral density matrix) associated to each SPOD mode for each frequency. The orange line denotes the sum of the spectral density of all 57 SPOD modes.}
\label{fig:SPOD_spectra_Thetap}
\end{figure}

Figure~\ref{fig:SPOD_spectra_Thetap} shows the SPOD spectra obtained for the symmetric and antisymmetric datasets for both jet separations, based on the cross-spectral density of the $\Theta$ fluctuations (see section \ref{sec:SPOD_method}). For all cases, the spectrum presents a broadband structure with highest energy contained in the range $St \approx [0.1,1]$, corresponding to the signature of the mixing noise. The decomposition is low-rank for most of the frequencies of interest, as indicated by the proximity of the line corresponding to mode 1 to the orange line, which represents the sum of the energy for all the modes. For frequencies below $St = 0.6$, more than 50\% of the total energy is contained in the first SPOD mode.

The structure of the first and second symmetric SPOD modes educed for $St = 0.4$ and $s/D = 1.76$ is shown in figures~\ref{fig:SPOD1_sym_vs_SS0_St0d4_sD176}$(a,c)$ and~\ref{fig:SPOD2_sym_vs_SS1_St0d4_sD176}$(a,c)$. Both the $\Theta$ fluctuation field as well as the schlieren fluctuation field obtained from the SPOD problem based on $\Theta$ ($\hat{\sigma}_{\Theta}$) are shown for each mode. SPOD mode 1 exhibits coherent structures that largely resemble toroidal fluctuations, while SPOD mode 2 features a clear flapping structure with low amplitude at the nozzle axes. The decomposition separates the two types of fluctuation into different SPOD modes, facilitating comparison with the PM-PSE models. The efficacy of the decomposition in separating toroidal and flapping modes is favoured by the fact that the two types of structure are naturally largely orthogonal to one another. In addition, both SPOD modes contain the associated Mach-wave radiation signature. These properties are observed over the entire frequency range of interest.

The peaks seen in the SPOD spectra at $St \approx 0.55$ and $St \approx 1.15$ are anomalies in the experimental data whose nature has not yet been clarified. Different tests have been conducted to shed light on their origin, in particular: schlieren measurements at different $M_j$, different jet spacing and different total temperature, as well as microphone measurements at different $M_j$. The results show that these peaks are not sensitive to the nozzle pressure ratio (or equivalently, to $M_j$) or the nozzle spacing, and they are consistently found in both schlieren and microphone measurements taken at different times. Their frequency, however, is found to be sensitive to the jet total temperature. At present, it is hypothesised that these anomalies are generated by structural vibrations in the twin-jet system, which weakly force the jets at those specific frequencies.

\subsection{Comparison between PM-PSE wavepackets and SPOD modes}

The comparison of PM-PSE wavepackets with the coherent structures educed from the schlieren measurements requires the calculation of a numerical schlieren field for the PM-PSE fluctuations. From the definition of the schlieren field measured in the experiments, the PM-PSE schlieren fluctuation field can be computed as

\begin{equation}
\hat{\sigma} (x,y) = \int_{-\infty}^{\infty} \frac{\partial \hat{\rho} (x,y,z)}{\partial x} \,\mathrm{d} z,
\end{equation}

\noindent where the density fluctuation is calculated from $\hat{p}$ and $\hat{T}$ through the linearized perfect gas equation of state, and the streamwise derivative of $\hat{\rho}$ is evaluated by differentiating the PM-PSE ansatz~\eqref{eq:PSE_ansatz} with respect to $x$. The PM-PSE $\hat{\Theta}$ fluctuation fields are then computed following the same procedure employed for the experimental datasets, i.e., solving the Poisson equation~\eqref{eq:Poisson_sch} in the spectral domain including a phase-speed filter for spurious waves that propagate with unphysically large supersonic speed~\citep{PadillaMontero2024}.

In order to quantify the structural similarity between the SPOD and PM-PSE modes, the following alignment factor (or projection coefficient) is calculated as the normalized projection of the schlieren SPOD eigenvectors on the corresponding PM-PSE schlieren perturbations for a given frequency, as done in previous studies of single jets~\citep{Gudmundsson:JFM11,Cavalieri:JFM13,Sinha:JFM14,Sasaki:JFM17}:

\begin{equation}
\chi_{\sigma_\Theta} = \frac{|\langle \hat{\sigma}_{\Theta}, \hat{\sigma}_{\textrm{PSE}} \rangle|}{\langle \hat{\sigma}_{\Theta}, \hat{\sigma}_{\Theta} \rangle^{1/2} \langle \hat{\sigma}_{\textrm{PSE}}, \hat{\sigma}_{\textrm{PSE}} \rangle^{1/2}},
\end{equation}

\noindent where the inner product between two fields $\langle \hat{q}_1, \hat{q}_2 \rangle$ is defined as

\begin{equation}
\langle \hat{q}_1, \hat{q}_2 \rangle = \iint_\Omega \hat{q}_1^* \hat{q}_2 \,\mathrm{d}x \mathrm{d}y,
\end{equation} 

\noindent with $\Omega$ denoting the spatial domain over which the projection is evaluated, in this case limited by the spatial window of the schlieren images. To avoid including the dark region corresponding to the mirror of the optical schlieren setup, a horizontal window between $x/D = 0.75$ and $x/D = 8.3$ is employed here for the calculation of the inner product. Alternatively to $\chi_{\sigma_\Theta}$, an alignment coefficient based on $\hat{\Theta}$ is also evaluated, given by

\begin{equation}
\chi_{\Theta} = \frac{|\langle \hat{\Theta}_{\textrm{SPOD}}, \hat{\Theta}_{\textrm{PSE}} \rangle|}{\langle \hat{\Theta}_{\textrm{SPOD}}, \hat{\Theta}_{\textrm{SPOD}} \rangle^{1/2} \langle \hat{\Theta}_{\textrm{PSE}}, \hat{\Theta}_{\textrm{PSE}} \rangle^{1/2}}.
\end{equation}

\subsubsection{Small jet separation, $s/D = 1.76$}
\label{sec:chi_sD176}

\begin{figure}
\centerline{\includegraphics[width=0.99\textwidth]{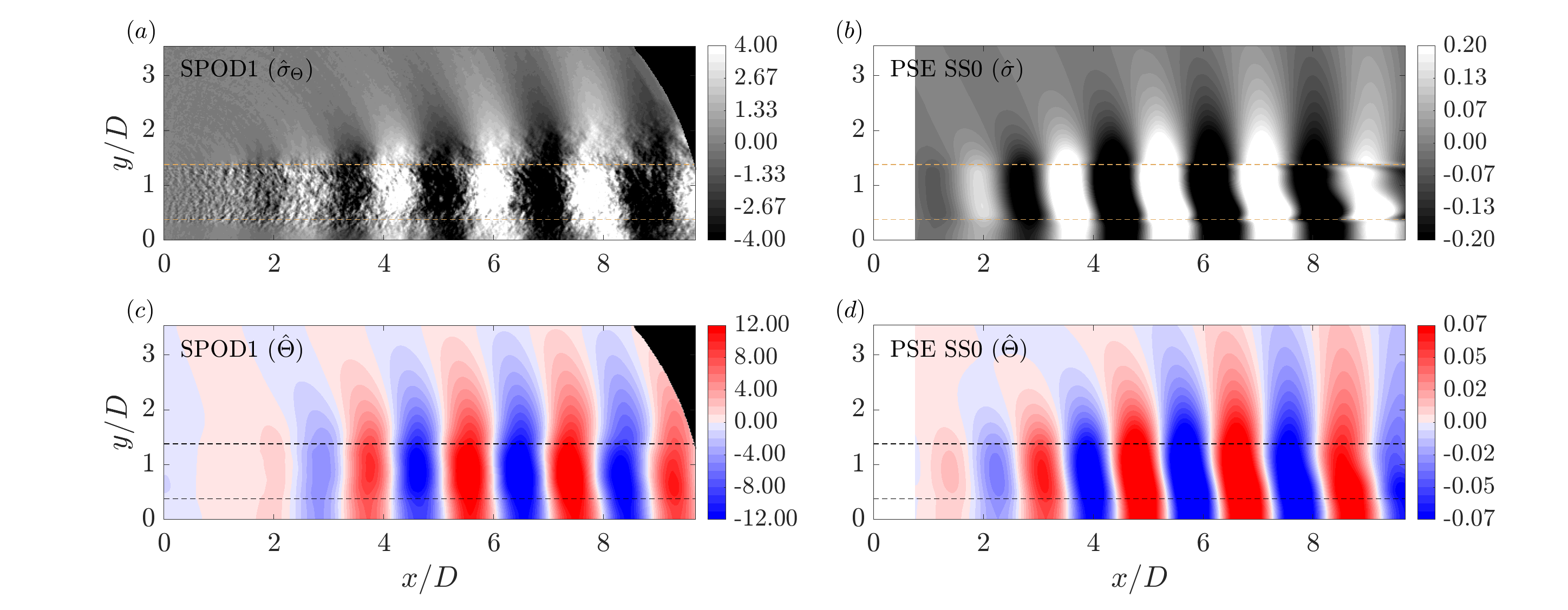}}
\caption{Contours of the real part of the symmetric toroidal fluctuation for $St = 0.4$ and $s/D = 1.76$: $(a)$ schlieren field of SPOD mode 1 ($\hat{\Theta}$-based CSD); $(b)$ schlieren field of PM-PSE mode SS0; $(c)$ $\hat{\Theta}$ field of SPOD mode 1; $(d)$ $\hat{\Theta}$ field of PM-PSE mode SS0. Dashed lines indicate the nozzle lip lines.}
\label{fig:SPOD1_sym_vs_SS0_St0d4_sD176}
\end{figure}

\begin{figure}
\centerline{\includegraphics[width=0.99\textwidth]{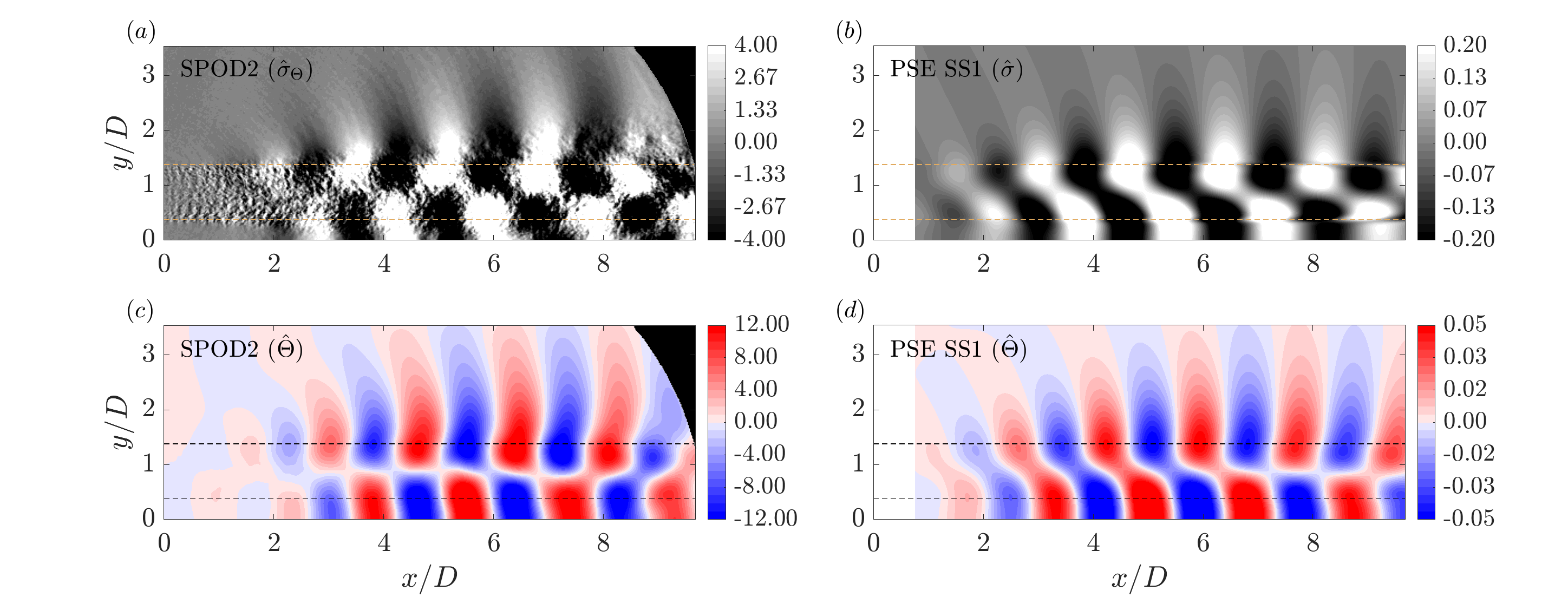}}
\caption{Contours of the real part of the symmetric flapping fluctuation for $St = 0.4$ and $s/D = 1.76$: $(a)$ schlieren field of SPOD mode 2; $(b)$ schlieren field of PM-PSE mode SS1; $(c)$ $\hat{\Theta}$ field of SPOD mode 2; $(d)$ $\hat{\Theta}$ field of PM-PSE mode SS1.}
\label{fig:SPOD2_sym_vs_SS1_St0d4_sD176}
\end{figure}

\begin{figure}
\centerline{\includegraphics[width=0.99\textwidth]{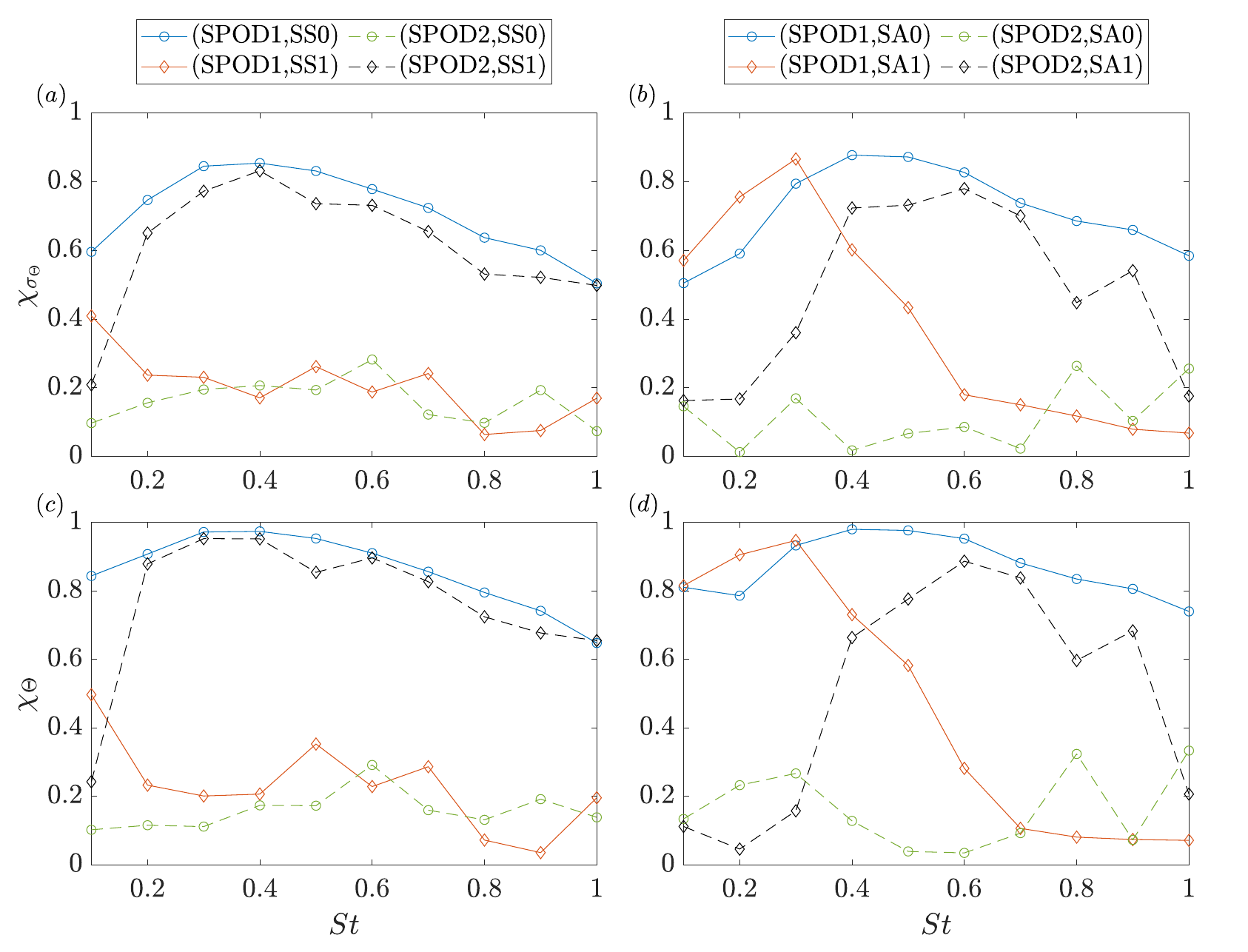}}
\caption{Alignment factors for $s/D = 1.76$: $(a,b)$ alignment between schlieren fluctuation fields; $(c,d)$ alignment between $\hat{\Theta}$ fields; $(a,c)$ symmetric fluctuations with respect to $xz$ plane; $(b,d)$ antisymmetric fluctuations with respect to $xz$ plane.}
\label{fig:alignment_sD176}
\end{figure}

First, the case of closely-spaced twin jets ($s/D = 1.76$) is discussed. A qualitative comparison between SPOD and PM-PSE fluctuations for symmetric modes is presented in figures~\ref{fig:SPOD1_sym_vs_SS0_St0d4_sD176} and \ref{fig:SPOD2_sym_vs_SS1_St0d4_sD176}. These figures show filled contour plots of the schlieren fluctuation field $\hat{\sigma}$ as well as the $\hat{\Theta}$ fluctuation field for both SPOD and PM-PSE modes at $St = 0.4$. Figure~\ref{fig:SPOD1_sym_vs_SS0_St0d4_sD176} compares SPOD mode 1 against the PM-PSE mode SS0, i.e., the symmetric toroidal fluctuation. The experimentally-educed structure and the modelled wavepacket are in excellent agreement, showing ring-like structures tilted with respect to the nozzle axis owing to the phase-speed mismatch between the inner and outer mixing layers, as discussed in section~\ref{sec:PSE_wavepackets}. The two representations of the fluctuation exhibit spatial wavenumber distributions that are in qualitative accordance both inside and outside of the jets, indicating that the Mach-wave radiation is also successfully captured by the PM-PSE model.

Figure~\ref{fig:SPOD2_sym_vs_SS1_St0d4_sD176} compares SPOD mode 2 with PM-PSE mode SS1. The results are also in good agreement in this case, with both representations showing symmetric flapping structures with similar wavenumbers and Mach-wave radiation patterns. The PM-PSE mode, however, does not yield a zero amplitude region at the nozzle axis. This is attributed to the phase-speed mismatch observed between the inner and outer mixing layers, which prevents the flapping mode structures from being perfectly in counter-phase with each other across the axis. In view of the results presented in section~\ref{sec:PSE_wavepackets}, this is considered a physical characteristic of closely-spaced twin jets. When performing the line-of-sight integration of the three-dimensional PM-PSE fluctuation field to obtain the associated schlieren field, the effect of such a mismatch is amplified. The SPOD algorithm, on the other hand, extracts a flapping structure which resembles more closely the $m = 1$ fluctuation encountered in single jets, featuring a checkerboard pattern with nearly zero amplitude on the nozzle axis. Since the SPOD is based on two-dimensional fields (schlieren images), no integration of a three-dimensional field along a spatial dimension is involved in this case, and the two-dimensional dynamics result in such a flapping motion. This discrepancy between the SPOD structure and the model, rather than implying a deficiency of PM-PSE in modelling flapping wavepackets, reflects a limitation of educing three-dimensional coherent structures from two-dimensional experimental images. The lack of information in the third dimension prevents SPOD from extracting a fluctuation that incorporates the effect of the line-of-sight integrated phase-speed mismatch.

Two alignment factors quantifying the agreement between PM-PSE and the experimentally-educed structures for $s/D = 1.76$ are shown in figure~\ref{fig:alignment_sD176}. Alignment factors are computed both for schlieren fluctuations (denoted by $\chi_{\sigma_\Theta}$) and for $\hat{\Theta}$ fluctuations (denoted by $\chi_\Theta$), as well as for symmetric and antisymmetric modes. For each case, four different alignments are evaluated: (i) first SPOD mode against the toroidal PM-PSE mode; (ii) second SPOD mode against the toroidal PM-PSE mode; (iii) first SPOD mode against the flapping PM-PSE mode; (iv) second SPOD mode against the flapping PM-PSE mode. For symmetric fluctuations, the alignment curves corresponding to each projection are labelled, respectively, as (SPOD1,SS0), (SPOD2,SS0), (SPOD1,SS1) and (SPOD2,SS1). According to the splitting of toroidal and flapping fluctuations into SPOD modes 1 and 2, respectively, alignments (SPOD1,SS0) and (SPOD2,SS1) are expected to yield much higher values than (SPOD2,SS0) and (SPOD1,SS1). This is what is observed in figure~\ref{fig:alignment_sD176}$(a,c)$, except for $St = 0.1$, for which the streamwise wavelength of the wavepacket structure is comparable to the size of the schlieren window and the resulting SPOD modes are not accurate representations. The alignment factors based on $\sigma_\Theta$ are significantly lower than those based on $\hat{\Theta}$ due to the presence of small-scale vortical structures in the fluctuation fields. The obtained alignment factors for both symmetric fluctuations are high over most of the studied frequency range. Particularly in the interval $St = [0.3, 0.5]$, values of $\chi_{\Theta} \approx 0.97$ ($\chi_{\sigma_\Theta} \approx 0.85$) are reached for toroidal fluctuations, and $\chi_{\Theta} \approx 0.95$ ($\chi_{\sigma_\Theta} \approx 0.83$) for flapping fluctuations, indicating an excellent agreement between the experimentally-educed coherent structures and the PM-PSE wavepacket models. According to the SPOD spectra shown in figure~\ref{fig:SPOD_spectra_Thetap}, this is the range of frequencies at which the coherent structures have highest energy.

\begin{figure}
\centerline{\includegraphics[width=0.99\textwidth]{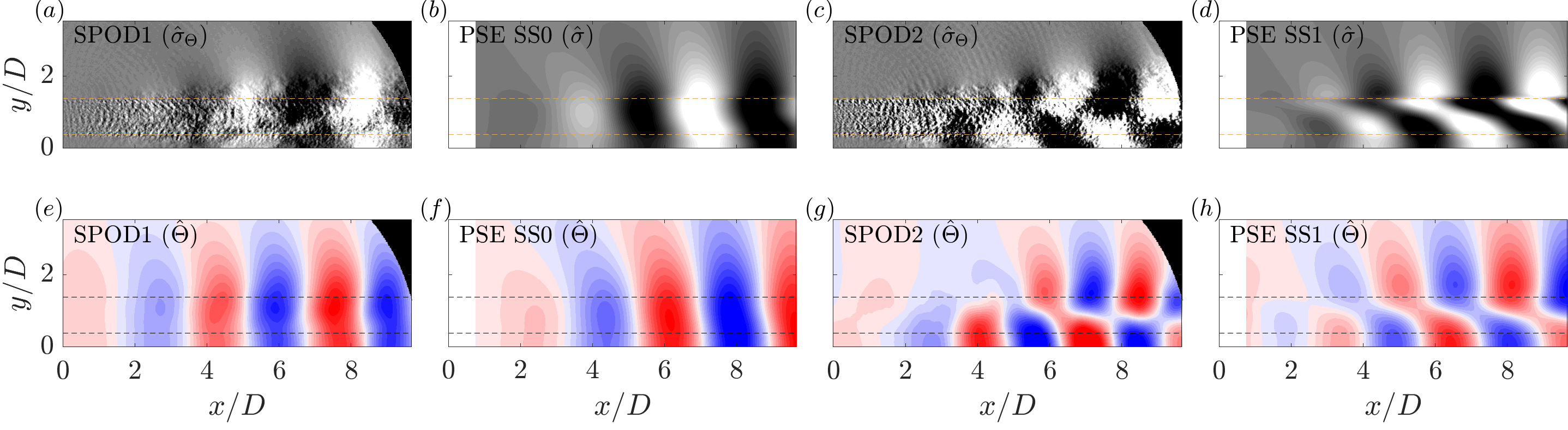}}
\caption{Contours of the real part of the symmetric fluctuations for $St = 0.2$ and $s/D = 1.76$: $(a)$ schlieren field of SPOD mode 1; $(b)$ schlieren field of PM-PSE mode SS0; $(c)$ schlieren field of SPOD mode 2; $(d)$ schlieren field of PM-PSE mode SS1; $(e)$ $\hat{\Theta}$ field of SPOD mode 1; $(f)$ $\hat{\Theta}$ field of PM-PSE mode SS0; $(g)$ $\hat{\Theta}$ field of SPOD mode 2; $(h)$ $\hat{\Theta}$ field of PM-PSE mode SS1.}
\label{fig:SPOD_sym_vs_SS_St0d2_sD176}
\end{figure}

\begin{figure}
\centerline{\includegraphics[width=0.99\textwidth]{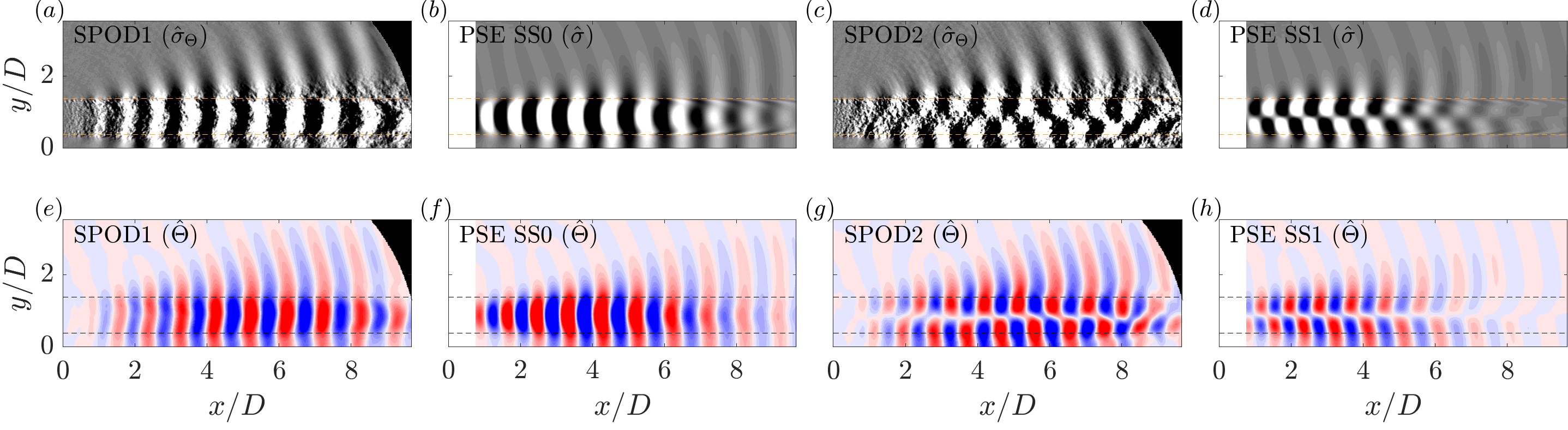}}
\caption{Contours of the real part of the symmetric fluctuations for $St = 0.8$ and $s/D = 1.76$: $(a)$ schlieren field of SPOD mode 1; $(b)$ schlieren field of PM-PSE mode SS0; $(c)$ schlieren field of SPOD mode 2; $(d)$ schlieren field of PM-PSE mode SS1; $(e)$ $\hat{\Theta}$ field of SPOD mode 1; $(f)$ $\hat{\Theta}$ field of PM-PSE mode SS0; $(g)$ $\hat{\Theta}$ field of SPOD mode 2; $(h)$ $\hat{\Theta}$ field of PM-PSE mode SS1.}
\label{fig:SPOD_sym_vs_SS_St0d8_sD176}
\end{figure}

For frequencies above $St = 0.5$ and below $St = 0.3$, the alignment factors (SPOD1,SS0) and (SPOD2,SS1) progressively decrease, although values above $\chi_{\sigma_\Theta} = 0.6$ are maintained throughout most of the studied frequency range. To illustrate the comparison between the symmetric wavepacket models and the experimentally-educed structures at other frequencies for which the alignment is lower, figures~\ref{fig:SPOD_sym_vs_SS_St0d2_sD176} and~\ref{fig:SPOD_sym_vs_SS_St0d8_sD176} show the comparison of symmetric SPOD modes and PM-PSE modes for $St = 0.2$ and $St = 0.8$, respectively. For low frequencies (e.g. $St = 0.2$), the deterioration of the agreement is attributed to the following observations: first, the streamwise wavelength of the PM-PSE wavepackets appears to be larger than that of the corresponding SPOD modes, especially for the toroidal fluctuations. This might reflect a limitation of the model at low frequencies. Past studies~\citep{Suzuki:JFM06,Gudmundsson:JFM11,Cavalieri:JFM13,Sinha:JFM14} have reported difficulties in modelling low-frequency wavepackets in single jets by means of PSE. Recent investigations based on resolvent analysis~\citep{Schmidt:JFM2018,Lesshafft:PRF2019,Pickering:JFM2020} have shown that non-modal effects are important for the linear dynamics at low $St$ and $m = 0$, and these cannot be captured by the PSE. Second, the streamwise range of linear growth and development of coherent structures for low $St$ is larger than for higher frequencies, which reach amplitude saturation earlier upstream. Given the limited streamwise length of the schlieren measurement window, the coherent fluctuations educed by SPOD at low $St$ exhibit an organized structure in a smaller portion of the domain. This is more evident for the symmetric flapping structure in SPOD2 at $St = 0.2$ (figure~\ref{fig:SPOD_sym_vs_SS_St0d2_sD176}$(c,g)$), which does not emerge as a clearly organized structure until $x/D > 5$. Third, the wavelength of the fluctuations for low $St$ becomes comparable to the size of the schlieren window. In this circumstance, the coherent structures extracted by means of SPOD are influenced by the domain truncation enforced by the size of the schlieren images. In particular, the size of the domain along $y$ limits the eduction of Mach-wave radiation after a certain streamwise position, which for e.g. $St = 0.2$, is found to be very early upstream ($x/D \approx 2$). The last two shortcomings are inherent to the methodology employed here for extracting coherent structures from the experimental data. In addition to the aforementioned limitations, the PM-PSE flapping modes for low $St$ are also highly distorted owing to the effect of the phase-speed mismatch described in section \ref{sec:PSE_wavepackets}.

For high frequencies ($St \geq 0.7$), the decrease in the alignment factors is mainly attributed to limitations of the linear PM-PSE model. As an illustrative example, for $St = 0.8$ (see figure~\ref{fig:SPOD_sym_vs_SS_St0d8_sD176}$(f,h)$) the PM-PSE wavepacket grows rapidly after the nozzle exit up to $x/D \approx 5$, and then progressively vanishes further downstream. The corresponding educed SPOD structures, in contrast, feature energetic coherent structures within the potential core extending up to the downstream boundary of the visualization window, without a decrease in the amplitude. This reflects a shortcoming of the linear PM-PSE formulation in modelling the behaviour downstream the linear growth region. Previous works on single jets have also observed similar discrepancies between the predictions of the linear stability approaches limited to modal instabilities and experimental measurements in the downstream region~\citep{Suzuki:JFM06,Gudmundsson:JFM11,Cavalieri:JFM13,Sinha:JFM14,Breakey:PRF2017}. Studies based on a spatial model of the Orr mechanism~\citep{Tissot:PRF2017}, spatial transient-growth calculations~\citep{Jordan:AIAAC2017} and resolvent analysis~\citep{Schmidt:JFM2018,Lesshafft:PRF2019} show that non-modal mechanisms become important in this region, and suggest that non-linear interactions are key in their activation~\citep{Jordan:AIAAC2017}. The PM-PSE wavepackets exhibit an additional difference: the amplitude of the Mach-wave radiation relative to the amplitude of the wavepacket evolving inside the jet is found to be weaker for the PM-PSE modes than for SPOD modes. This discrepancy has also been observed in previous studies dealing with isolated jets~\citep{Sinha:JFM14,Breakey:PRF2017} and has been attributed to inherent limitations of the parabolized stability equations~\citep{Towne:TCFD2019}.

The aforementioned effects have a small impact for the intermediate frequency range, $St \approx [0.3, 0.5]$, which corresponds to the most energetic part of the spectrum according to the experimental data (see figure~\ref{fig:SPOD_spectra_Thetap}). Large alignment factors between the PM-PSE model and the experimentally-educed structures are obtained for these frequencies. This result is consistent with the dominance of modal, Kelvin-Helmholtz instabilities in the intermediate frequency range highlighted by~\citet{Pickering:JFM2020} for single jets.

\begin{figure}
\centerline{\includegraphics[width=0.99\textwidth]{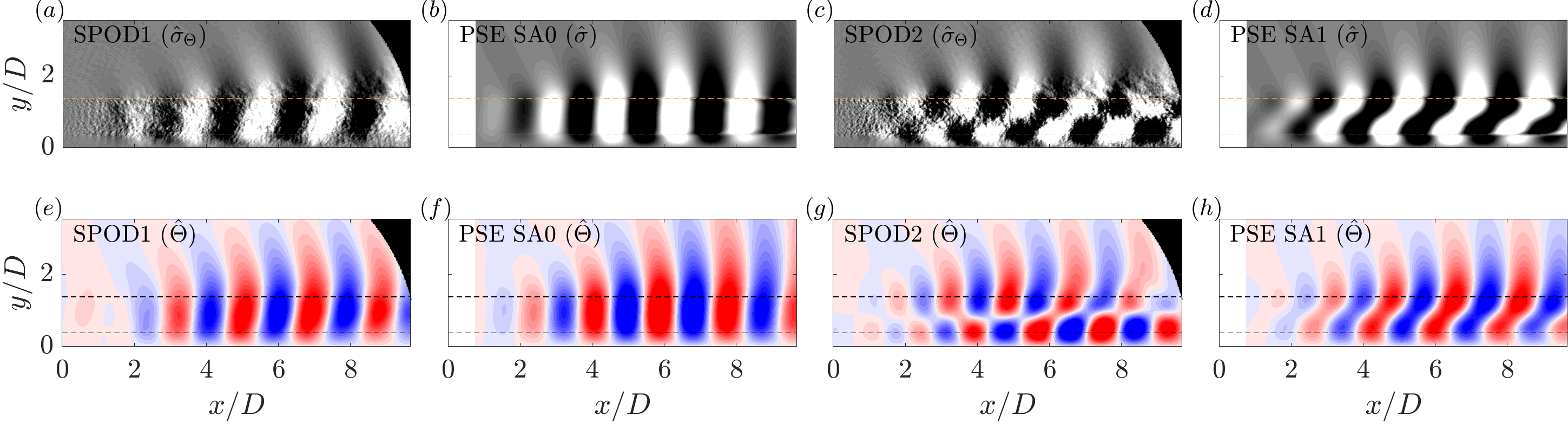}}
\caption{Contours of the real part of the antisymmetric fluctuations for $St = 0.4$ and $s/D = 1.76$: $(a)$ schlieren field of SPOD mode 1; $(b)$ schlieren field of PM-PSE mode SA0; $(c)$ schlieren field of SPOD mode 2; $(d)$ schlieren field of PM-PSE mode SA1; $(e)$ $\hat{\Theta}$ field of SPOD mode 1; $(f)$ $\hat{\Theta}$ field of PM-PSE mode SA0; $(g)$ $\hat{\Theta}$ field of SPOD mode 2; $(h)$ $\hat{\Theta}$ field of PM-PSE mode SA1.}
\label{fig:SPOD_asym_vs_SA_St0d4_sD176}
\end{figure}

\begin{figure}
\centerline{\includegraphics[width=0.99\textwidth]{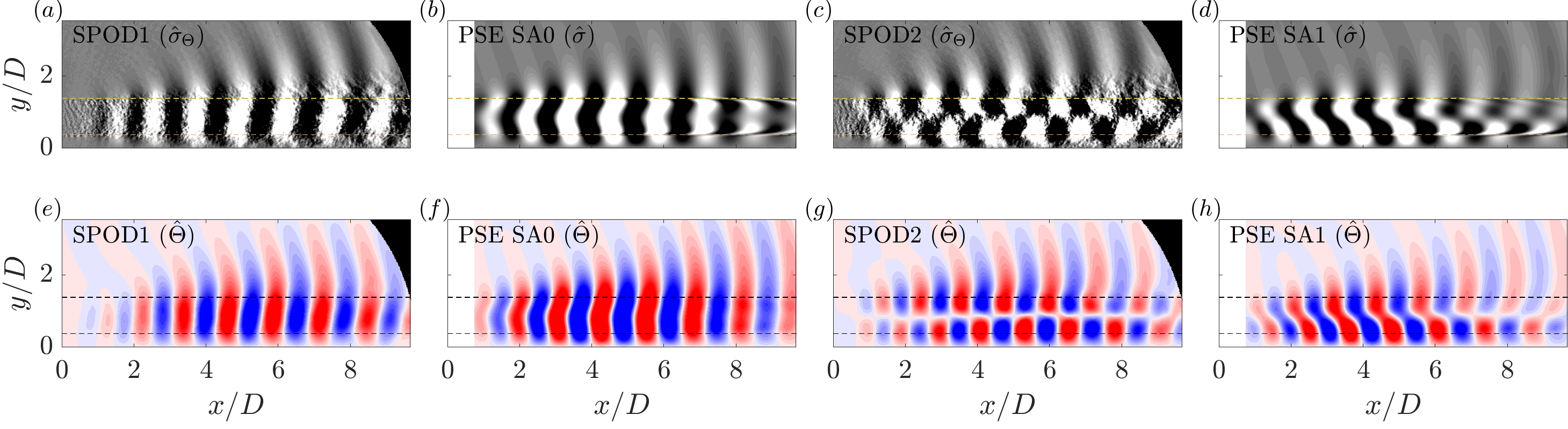}}
\caption{Contours of the real part of the antisymmetric fluctuations for $St = 0.6$ and $s/D = 1.76$: $(a)$ schlieren field of SPOD mode 1; $(b)$ schlieren field of PM-PSE mode SA0; $(c)$ schlieren field of SPOD mode 2; $(d)$ schlieren field of PM-PSE mode SA1; $(e)$ $\hat{\Theta}$ field of SPOD mode 1; $(f)$ $\hat{\Theta}$ field of PM-PSE mode SA0; $(g)$ $\hat{\Theta}$ field of SPOD mode 2; $(h)$ $\hat{\Theta}$ field of PM-PSE mode SA1.}
\label{fig:SPOD_asym_vs_SA_St0d6_sD176}
\end{figure}

The alignment factors for the antisymmetric fluctuations are shown in figure~\ref{fig:alignment_sD176}$(b,d)$. A clear dominance of alignment factors (SPOD1,SA0) and (SPOD2,SA1) is also observed (as for the symmetric structures) for frequencies above $St = 0.4$. An anomalous behaviour is found at lower frequencies for which both PM-PSE modes SA0 and SA1 have a high alignment against SPOD mode 1. This is discussed in more detail in the next section (\ref{sec:inconsistent_chi}). Figures~\ref{fig:SPOD_asym_vs_SA_St0d4_sD176} and \ref{fig:SPOD_asym_vs_SA_St0d6_sD176} compare the first two antisymmetric SPOD modes against the PM-PSE modes SA0 and SA1 for $St = 0.4$ and $St = 0.6$, respectively. For toroidal fluctuations, the comparison at these frequencies reveals the same findings as for the symmetric case, namely, a good agreement for mid-range frequencies ($St = 0.4$), which progressively deteriorates for higher and lower frequencies. For antisymmetric flapping fluctuations, their distortion, owing to the phase-speed mismatch between the inner and outer mixing layers, is found to be stronger than for the symmetric counterpart. This is supported by figure~\ref{fig:phase_diff_vs_x_St0d4}$(a)$, where mode SA1 has the highest accumulated phase difference. This distortion reduces the agreement with SPOD modes, and shifts the region of optimal alignment towards higher frequencies, namely, $St = 0.6$ instead of $St = 0.4$ in the symmetric case.

\subsubsection{On the inconsistent agreement for antisymmetric flapping modes at low $St$} \label{sec:inconsistent_chi}

\begin{figure}
\centerline{\includegraphics[width=0.99\textwidth]{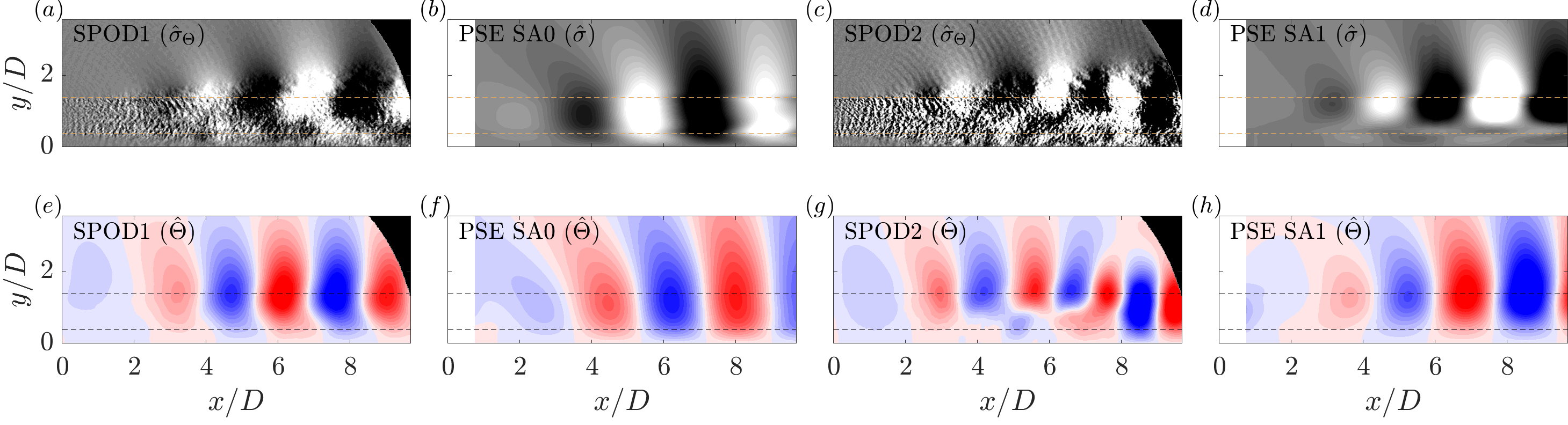}}
\caption{Contours of the real part of the antisymmetric fluctuations for $St = 0.2$ and $s/D = 1.76$: $(a)$ schlieren field of SPOD mode 1; $(b)$ schlieren field of PM-PSE mode SA0; $(c)$ schlieren field of SPOD mode 2; $(d)$ schlieren field of PM-PSE mode SA1; $(e)$ $\hat{\Theta}$ field of SPOD mode 1; $(f)$ $\hat{\Theta}$ field of PM-PSE mode SA0; $(g)$ $\hat{\Theta}$ field of SPOD mode 2; $(h)$ $\hat{\Theta}$ field of PM-PSE mode SA1.}
\label{fig:SPOD_asym_vs_SA_St0d2_sD176}
\end{figure}

\begin{figure}
\centerline{\includegraphics[width=0.99\textwidth]{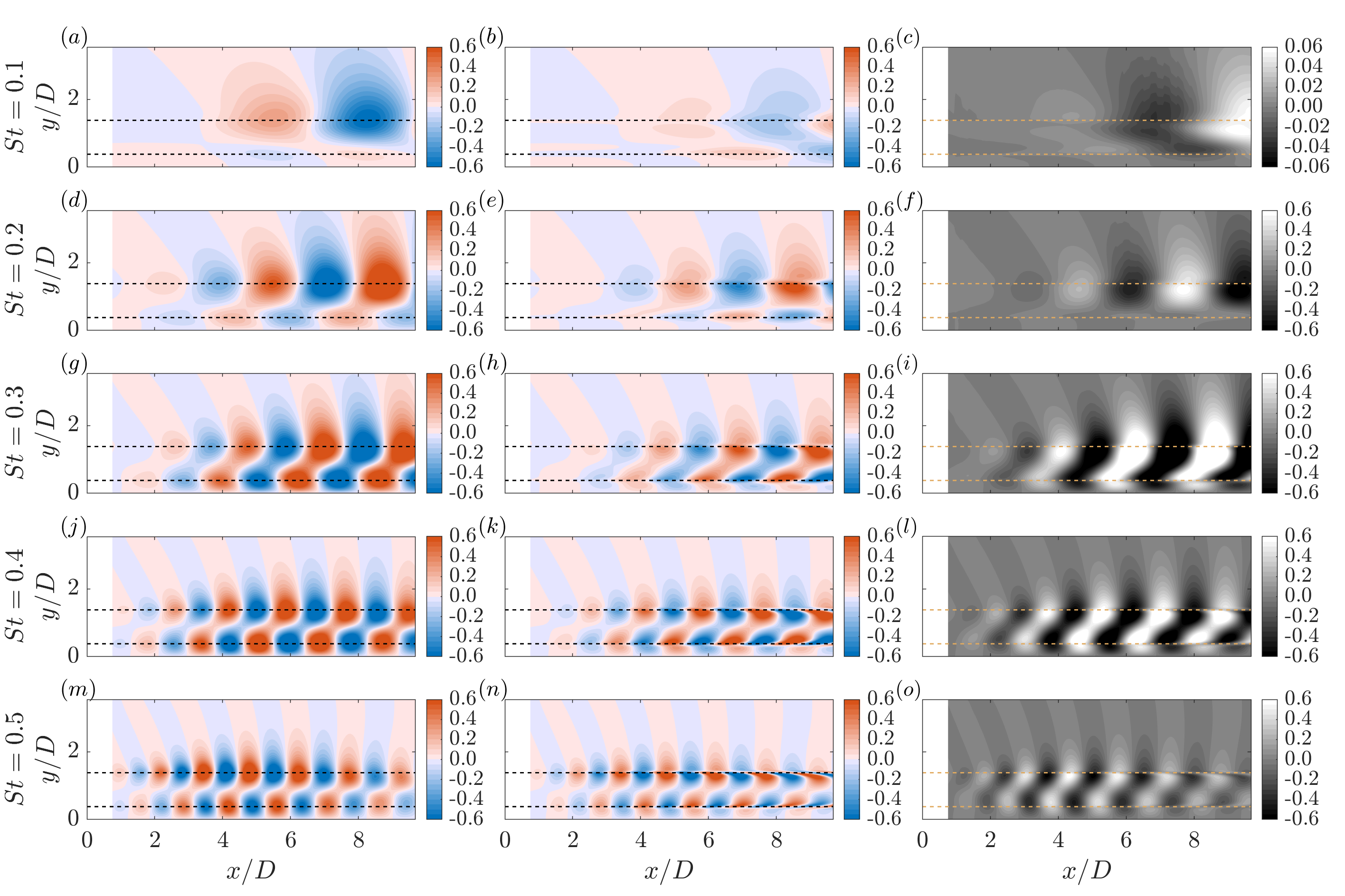}}
\caption{Contours of the real part of the PM-PSE antisymmetric flapping fluctuations (SA1) for various $St$ ($s/D = 1.76$): $(a,d,g,j,m)$ pressure fluctuation at $z = 0$; $(b,e,h,k,n)$ density fluctuation at $z = 0$; $(c,f,i,l,o)$ schlieren fluctuation; $(a,b,c)$ $St = 0.1$; $(d,e,f)$ $St = 0.2$; $(g,h,i)$ $St = 0.3$; $(j,k,l)$ $St = 0.4$; $(m,n,o)$ $St = 0.5$.}
\label{fig:PSE_SA1_vs_St}
\end{figure}

The alignment factor for antisymmetric flapping modes at low frequencies ($St < 0.4$, see figure~\ref{fig:alignment_sD176}$(b,d)$) shows an inconsistent behaviour in which both PM-PSE modes SA0 and SA1 have a high alignment with SPOD mode 1. To illustrate this, figure~\ref{fig:SPOD_asym_vs_SA_St0d2_sD176} compares empirical (SPOD) and modelled (PM-PSE) antisymmetric fluctuations for $St = 0.2$. The PM-PSE schlieren fluctuation and the $\Theta$ fluctuation fields in this case better resemble a toroidal fluctuation than a flapping one. The structure of the line-of-sight integrated fields for SA1 at this frequency is similar to that of mode SA0 and to that of mode SPOD1. As a result, both the (SPOD1,SA0) and (SPOD1,SA1) factors yield similar alignments. In addition, the educed coherent structure in SPOD mode 2 is poorly organized for most of the measurement window at this frequency, leading to small values of $\chi_{\sigma_\Theta}$ for the projection (SPOD2,SA1).

The explanation of this phenomenon is linked to the deformation suffered by the antisymmetric flapping fluctuations when the jets are closely spaced and the value of $St$ is low. This deformation is, on one side, caused by the phase-speed mismatch induced by the mean-flow differences between the inner and outer mixing layers of the twin-jet system, as described in previous sections. On the other side, flapping antisymmetric modes modelled by PM-PSE are confined between the nozzle axis and the symmetry plane at $y = 0$ (where antisymmetry boundary conditions are imposed). Under the combination of low $St$ and small jet spacing, the resulting Kelvin-Helmholtz wavelength is large compared with the distance between the nozzle axis and the symmetry axis at $y = 0$. As a consequence, the flapping wavepacket structures for $s/(2D) > y/D > 0$ become squeezed in the vertical direction and elongated in the streamwise direction, leading to a phase-speed increase in this region and to fluctuations which, after line-of-sight integration, adopt a shape such as the one depicted in figure~\ref{fig:SPOD_asym_vs_SA_St0d2_sD176}$(d)$. To support this argument, figure~\ref{fig:PSE_SA1_vs_St} represents the evolution of PM-PSE mode SA1 as a function of $St$. This figure shows contours of the PM-PSE pressure fluctuation, density fluctuation and schlieren fluctuation for $St = [0.1,0.5]$ in the $xy$ plane at $z = 0$. As $St$ is progressively decreased, the wavelength of the wavepackets becomes larger, the phase shift between the flapping structures above and below the nozzle axis deviates from the value of $\upi$ radians for an exact $m=1$, and the shape of the structures below the nozzle axis progressively becomes more distorted. When looking at schlieren PSE fluctuations, the checkerboard pattern characteristic of pure flapping motions progressively fades as $St$ decreases and the line-of-sight integrated wavepacket signature shifts to structures that appear closer to toroidal motions tilted at an angle with respect to the $y$ axis. On the other hand, although the SPOD is able to decompose the experimental data into structures that strongly resemble toroidal and flapping fluctuations in the leading modes, there is no guarantee that the decomposition would strictly separate the structures into the actual, $m$-like twin-jet modes (SA0, SA1, SA2 etc.). Instead, it tends to build modes that are spatially orthogonal within the defined inner product. In consequence, the individual SPOD modes may fail to capture the subtleties of the phase-speed mismatch and shape distortion observed in the toroidal and flapping antisymmetric PM-PSE modes, features which deteriorate the orthogonality and that, in turn, may be spread over different SPOD modes. This limitation associated with the use of schlieren (two-dimensional) visualizations to educe the empirical structures leads to the anomalous behaviour in the alignment factors reported in figure~\ref{fig:alignment_sD176}$(b,d)$.

\subsubsection{Moderate jet separation, $s/D = 3$}

\begin{figure}
\centerline{\includegraphics[width=0.99\textwidth]{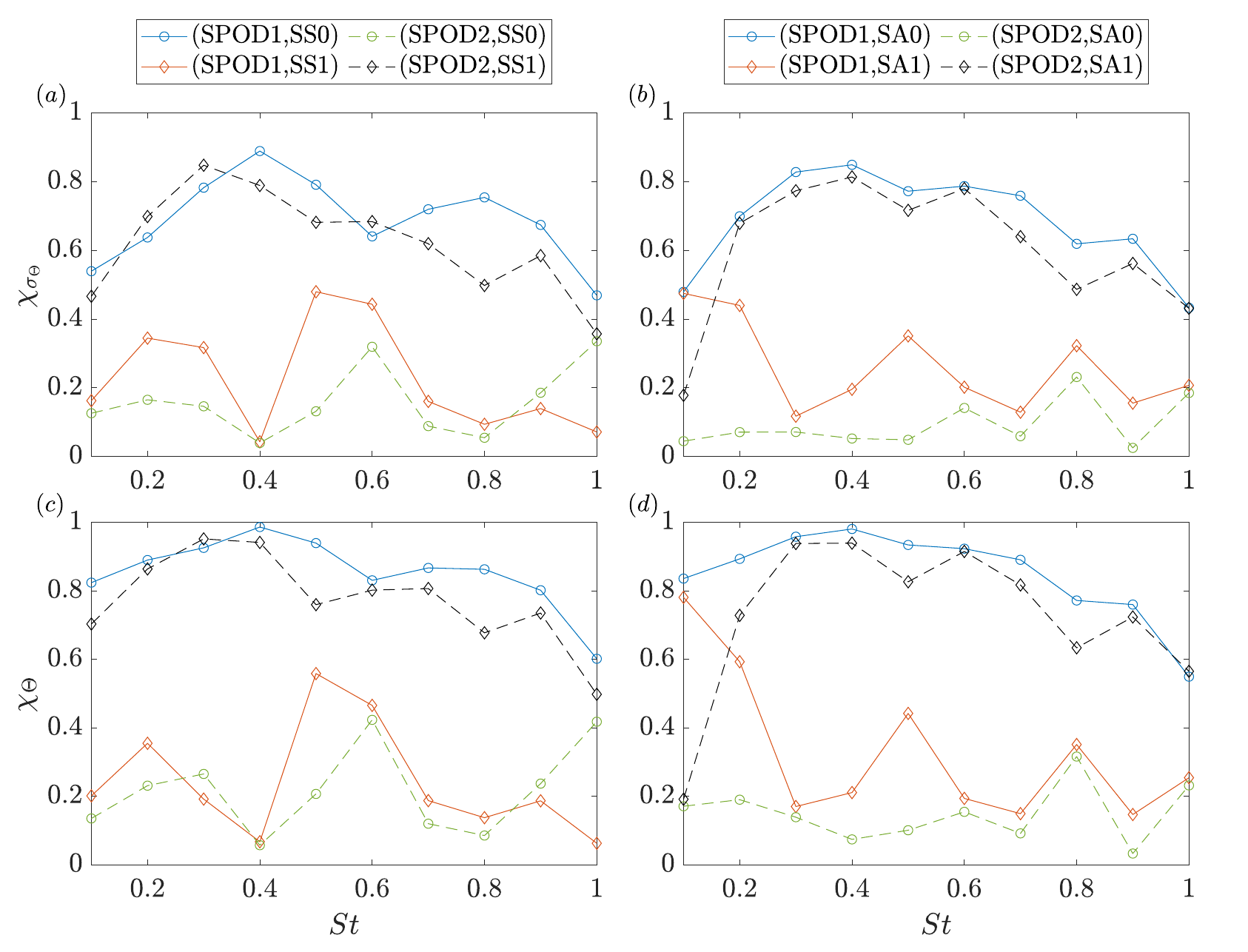}}
\caption{Alignment factors for $s/D = 3$: $(a,b)$ alignment between schlieren fluctuation fields; $(c,d)$ alignment between $\hat{\Theta}$ fields; $(a,c)$ symmetric fluctuations with respect to $xz$ plane; $(b,d)$ antisymmetric fluctuations with respect to $xz$ plane.}
\label{fig:alignment_sD3}
\end{figure}

\begin{figure}
\centerline{\includegraphics[width=0.99\textwidth]{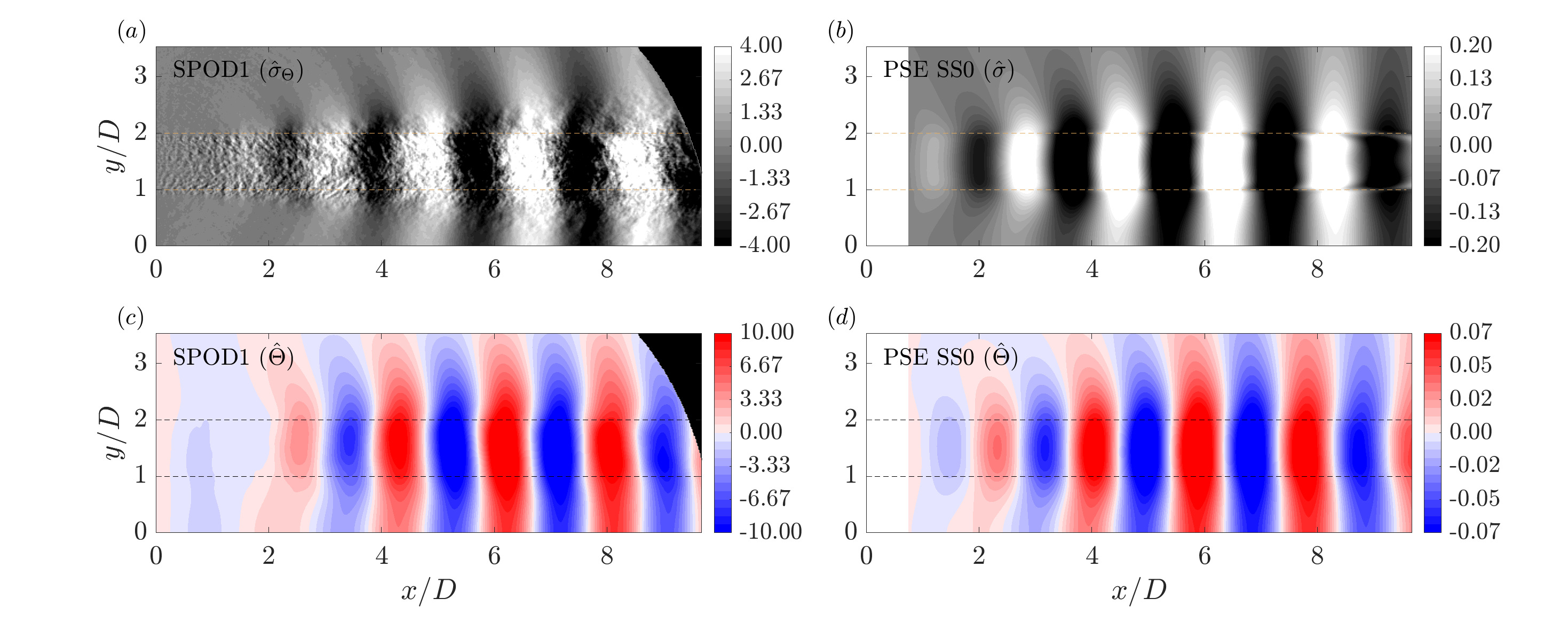}}
\caption{Contours of the real part of the symmetric toroidal fluctuation for $St = 0.4$ and $s/D = 3$: $(a)$ schlieren field of SPOD mode 1; $(b)$ schlieren field of PM-PSE mode SS0; $(c)$ $\hat{\Theta}$ field of SPOD mode 1; $(d)$ $\hat{\Theta}$ field of PM-PSE mode SS0.}
\label{fig:SPOD1_sym_vs_SS0_St0d4_sD3}
\end{figure}

\begin{figure}
\centerline{\includegraphics[width=0.99\textwidth]{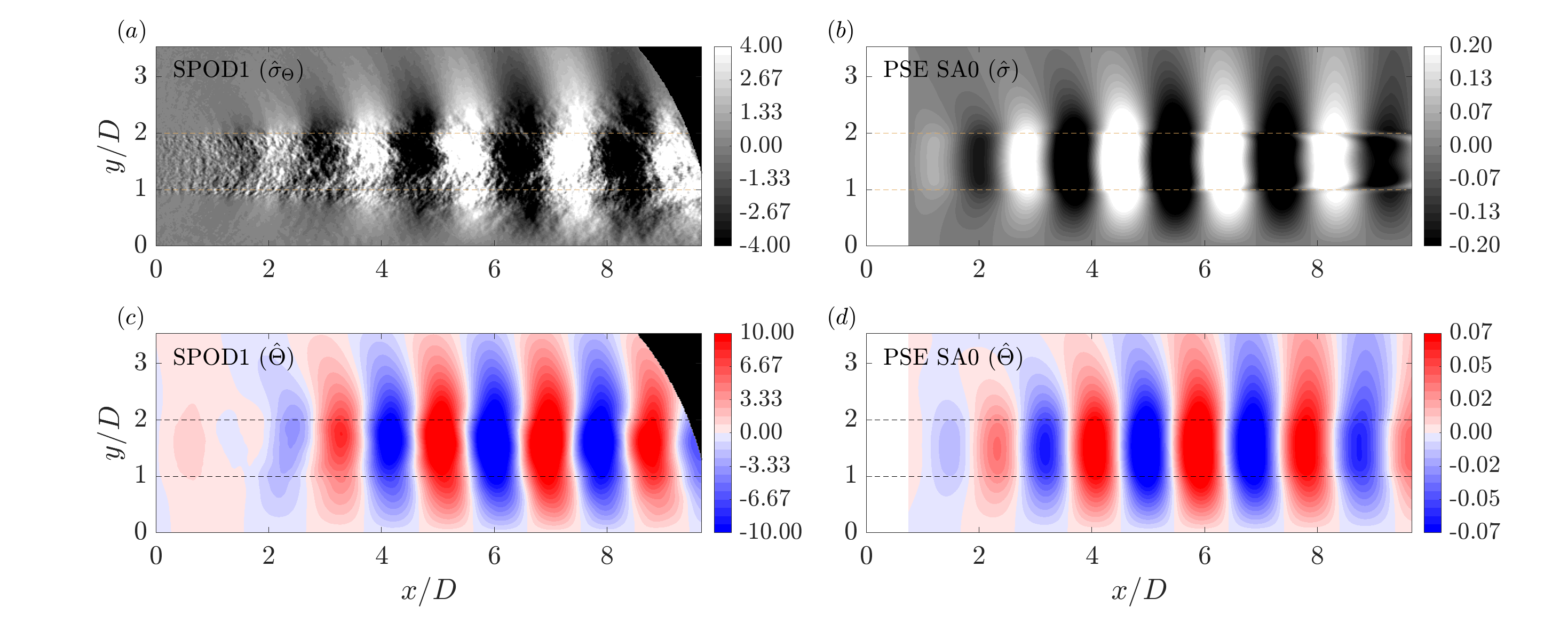}}
\caption{Contours of the real part of the antisymmetric toroidal fluctuation for $St = 0.4$ and $s/D = 3$: $(a)$ schlieren field of SPOD mode 1; $(b)$ schlieren field of PM-PSE mode SA0; $(c)$ $\hat{\Theta}$ field of SPOD mode 1; $(d)$ $\hat{\Theta}$ field of PM-PSE mode SA0.}
\label{fig:SPOD1_asym_vs_SA0_St0d4_sD3}
\end{figure}

\begin{figure}
\centerline{\includegraphics[width=0.99\textwidth]{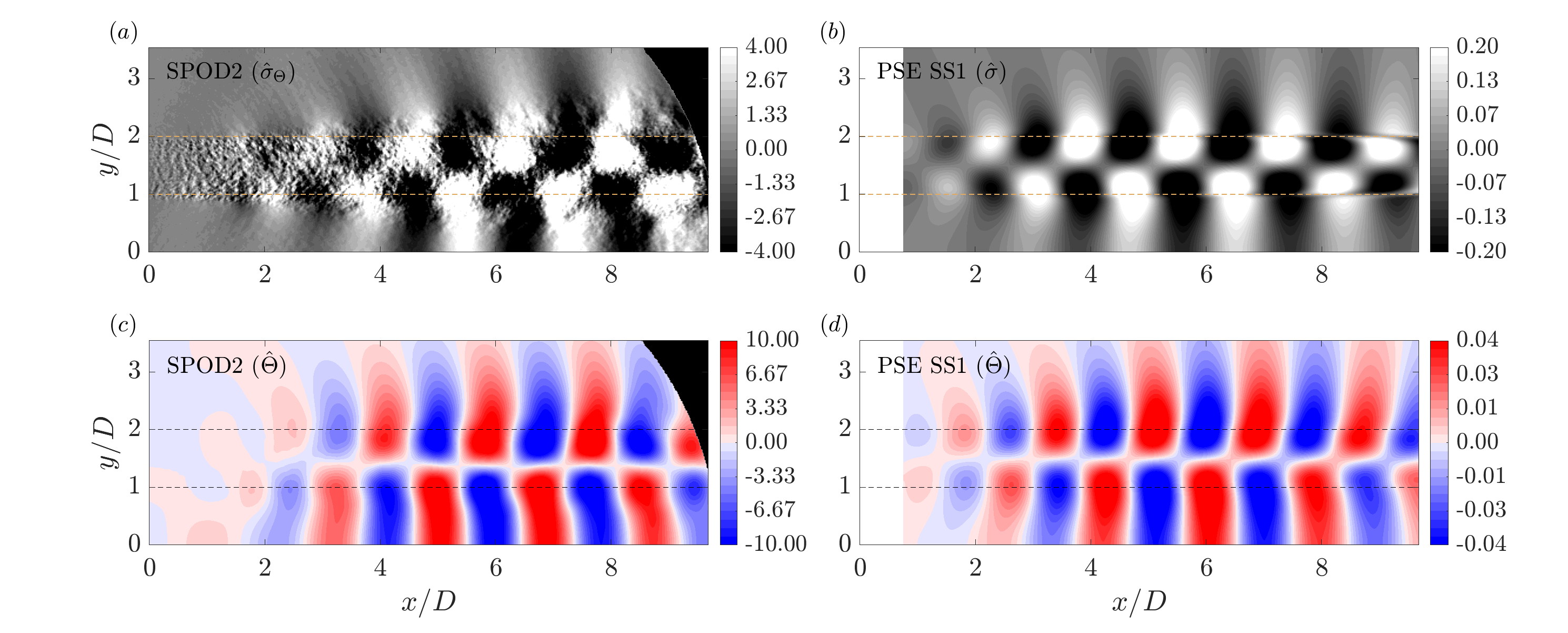}}
\caption{Contours of the real part of the symmetric flapping fluctuation for $St = 0.4$ and $s/D = 3$: $(a)$ schlieren field of SPOD mode 2; $(b)$ schlieren field of PM-PSE mode SS1; $(c)$ $\hat{\Theta}$ field of SPOD mode 2; $(d)$ $\hat{\Theta}$ field of PM-PSE mode SS1.}
\label{fig:SPOD2_sym_vs_SS1_St0d4_sD3}
\end{figure}

\begin{figure}
\centerline{\includegraphics[width=0.99\textwidth]{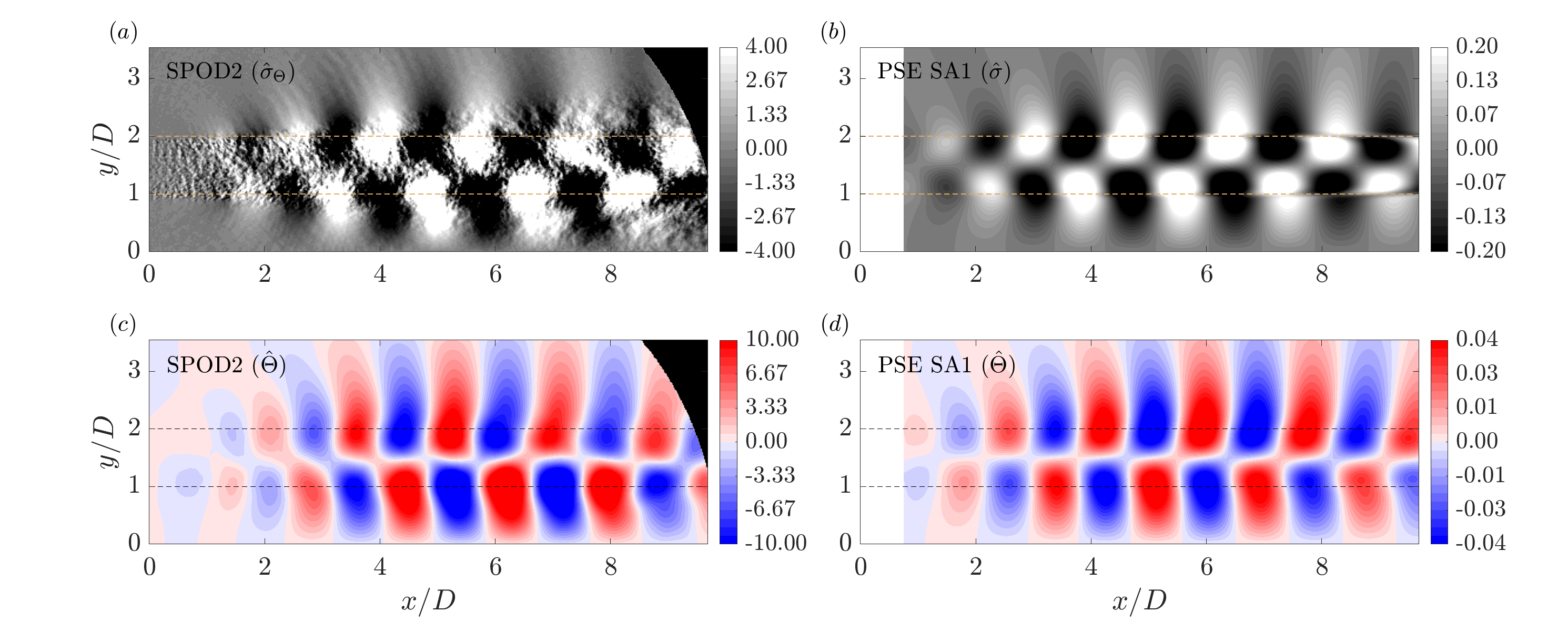}}
\caption{Contours of the real part of the antisymmetric flapping fluctuation for $St = 0.4$ and $s/D = 3$: $(a)$ schlieren field of SPOD mode 2; $(b)$ schlieren field of PM-PSE mode SA1; $(c)$ $\hat{\Theta}$ field of SPOD mode 2; $(d)$ $\hat{\Theta}$ field of PM-PSE mode SA1.}
\label{fig:SPOD2_asym_vs_SA1_St0d4_sD3}
\end{figure}

The comparison between SPOD and PM-PSE for $s/D = 3$ is reported in figures~\ref{fig:alignment_sD3} to~\ref{fig:SPOD2_asym_vs_SA1_St0d4_sD3}. Figure~\ref{fig:alignment_sD3} presents the alignment factors, which show good agreement between SPOD1 and toroidal PM-PSE modes, and between SPOD2 and flapping PM-PSE modes. For this jet spacing, the anomaly in the alignment induced by the deformation of antisymmetric flapping modes manifests only at very low frequencies ($St < 0.2$), which is consistent with the fact that the larger jet spacing does not constrain the development of antisymmetric flapping structures over a longer range of wavelengths. Similarly, since the phase-speed mismatch between the inner and outer mixing layers is much smaller for this nozzle spacing (see figures~\ref{fig:phase_diff_vs_x_St0d4}$(b)$ and~\ref{fig:cph_vs_x_St0d4}$(e$-$h)$), the distortion of symmetric flapping structures is also considerably weaker, and the evolution of the alignment factor as a function of $St$ is close to the evolution of the alignment for toroidal modes. These observations are supported by figures~\ref{fig:SPOD1_sym_vs_SS0_St0d4_sD3} and ~\ref{fig:SPOD1_asym_vs_SA0_St0d4_sD3}, which show the symmetric and antisymmetric comparisons for toroidal fluctuations at $St = 0.4$, as well as figures~\ref{fig:SPOD2_sym_vs_SS1_St0d4_sD3} and~\ref{fig:SPOD2_asym_vs_SA1_St0d4_sD3}, which display the respective comparison for flapping fluctuations. For $s/D=3$, the toroidal structures are almost perpendicular to the nozzle axes, and the modelled flapping wavepackets present a well-defined checkerboard pattern, respectively resembling the $m = 0$ and $m = 1$ modes found in isolated round jets. This reflects how, for $s/D = 3$, the interaction between jets is much weaker than for $s/D = 1.76$, with each jet behaving closer to an isolated axisymmetric case.

The magnitude of the alignment factors for $s/D = 3$ is not significantly higher than for $s/D = 1.76$, especially near the best alignment frequency $St \approx 0.4$. This is a quantitative indicator that the PM-PSE wavepacket models for $s/D = 1.76$ are as accurate as for $s/D = 3$, and therefore are able to incorporate the 
impact of the mean-flow interactions occurring when the jets are closely spaced.

The alignment factors obtained for both values of $s/D$ are comparable to those reported in previous studies dealing with single round jets (see for example~\citet{Gudmundsson:JFM11,Cavalieri:JFM13,Sinha:JFM14}), demonstrating the ability of plane-marching PSE to successfully model wavepackets in perfectly-expanded supersonic twin jets for a significant portion of the frequency spectrum, and especially for the most energetic part of the spectrum according to SPOD.

\section{Conclusions}
\label{sec:conclusions}

The generation of mixing noise by isolated supersonic turbulent jets is known to be related to coherent structures (wavepackets) which can be successfully described as Kelvin-Helmholtz instabilities supported by the mean flow~\citep{CrightonGaster:JFM76,TamBurton1984b,YenMessersmithAIAA1999,Piot:IJAAF06,Rodriguez2013,Sinha:JFM14}.  The performance of mean-flow-based linear stability calculations in describing coherent structures in supersonic twin jets, however, has not yet been fully characterized. Models are available based on simplified descriptions of the three-dimensional mean flow (see for instance~\citet{Rodriguez2021}), which have been successful in computing wavepacket structures but which lack experimental validation.

This work contributes to the modelling of twin-jet coherent structures, important for mixing noise, in two ways: by providing a higher fidelity, yet affordable description of the twin-jet dynamics using three-dimensional RANS calculations combined with plane-marching parabolized stability equations, and by providing the first quantitative validation of wavepacket models based on linear stability theory against experimental measurements of supersonic twin jets.

The analysis focused on supersonic round twin-jets generated by convergent-divergent nozzles operating at perfectly-expanded conditions. Two different nozzle spacings $s/D$ were considered, consisting of a closely-spaced case ($s/D = 1.76$) featuring a strong interaction between jets, and a moderately-spaced configuration ($s/D = 3$), characterized by a weaker interaction between the jets. Experimental measurements were performed to provide validation data. In particular, PIV measurements of the mean velocity field were made to compare against the RANS calculations, allowing quantitative comparisons in the symmetry plane containing the two jets ($y/D = 0$), and high-speed schlieren visualizations were performed to extract coherent structures present in the twin-jet system by means of SPOD, enabling qualitative and quantitative comparison against the PM-PSE fluctuations.

The three-dimensional RANS calculations were performed using a second-order finite volume solver featuring the Menter SST turbulence model. Modified turbulence model constants were employed, calibrated by linear interpolation between the original values provided by~\citet{Menter1994} and the optimized values provided by~\citet{Ozawa2024} for a single supersonic jet. The resulting mean-flow solutions are found to be in good agreement with the PIV measurements, demonstrating the ability of the RANS model to properly account for the non-linear mean-flow interaction between the jets. The RANS results illustrate important mean-flow differences that emerge between the inner and outer mixing layers of each jet when the nozzle separation is small, such as for $s/D = 1.76$.

Plane-marching PSE wavepackets were computed for a range of frequencies relevant for mixing noise ($St = [0.1,1]$). The computed structures exhibit toroidal and flapping fluctuations following the natural symmetries of the twin-jet system, together with a non-axisymmetric Mach-wave radiation signature. 
Owing to the schlieren setup, only those fluctuations symmetric with respect to $z = 0$ can be visualized. The analysis was thus focused on toroidal and flapping fluctuations symmetric with respect to $z = 0$, namely oscillation modes SS0, SS1, SA0 and SA1. The PM-PSE calculations for closely-spaced twin jets reveal significant differences in the phase of structures between the inner and the outer shear layers, especially for lower frequencies ($St \leq 0.4$), where interaction between the two jets is strong in terms of both non-linear (mean-flow) and linear (wavepacket) dynamics. For symmetric modes, these differences are found to be induced by a mismatch in the phase speed of the instabilities which correlates with differences in the streamwise mean flow velocity. For antisymmetric modes, in addition to the mismatch induced by differences in mean-flow velocity, changes in the phase speed are also caused by a spatial constraint of the wavepacket structures in the shear layers that develop between the two jets. With some limitations inherent to the PSE formulation~\citep{Towne:TCFD2019}, the modelled wavepackets for supersonic jets directly capture their Mach-wave radiation, which dominates the near-field structure.

A methodology for obtaining comparisons of the PM-PSE wavepacket models against experimentally-educed structures is also presented, making use of a recently developed approach that facilitates the extraction of coherent structures from schlieren images~\citep{Prasad2022,PadillaMontero2024}. The technique derives the line-of-sight integrated streamwise derivative of the momentum potential field from the schlieren images and extracts coherent structures of the momentum potential field instead of the original schlieren field, by means of SPOD. Comparison of SPOD modes and PM-PSE wavepackets reveals good qualitative agreement for symmetric and antisymmetric toroidal structures, as well as for symmetric flapping structures for the frequency range considered. Quantitative comparisons are performed through the definition of an alignment coefficient which measures the normalized projection of one field into the other. The alignment factors show agreement to be especially high in the range of frequencies at which the energy of the SPOD spectra is maximum ($St = [0.3,0.5]$), providing a first validation of the modelled wavepacket properties.

The results presented herein show that the modelling strategy based on RANS and PM-PSE can be useful for the physical understanding of how the interaction between the two jets affects the wavepacket properties, and in particular, for parametric studies exploring different jet spacings and nozzle shapes. Agreement with the experimental measurements confirms mean-flow-based linear stability calculations can correctly capture coherent structures present in the twin-jet flow at conditions where non-modal effects are not important. More sophisticated approaches based on linear stability equations which enable the recovery of non-modal instabilities, like resolvent analysis \citep{Garnaud:JFM2013,Jeun:PF2016,Schmidt:JFM2018} or one-way stability equations \citep{Towne:JCP15} can offer improved modelling capabilities over the PM-PSE employed here, but their application to fully three-dimensional flows like the twin-jet configuration remains more computationally expensive despite recent developments~\citep{Martini:JFM2021,Towne:JFM2022}. In scenarios involving shape optimisation via iterative processes, where reducing the computational cost is of primary importance, the PSE approach continues to be of high interest.

New sets of experimental data remain necessary for the validation of structures antisymmetric with respect to $z = 0$, as well as for assessing the accuracy of the growth rates predicted by the linear theory. Finally, the methodology presented is only strictly valid for perfectly-expanded conditions. Coherent structures developing in supersonic twin jets operating at over- or underexpanded conditions may be dominated by screech resonances, which pose additional modelling challenges for linear stability theory owing to the strong mean-flow gradients induced by shock and expansion waves, and the presence of upstream travelling waves~\citep{Edgington-Mitchell2022}.

\backsection[Acknowledgements]{The authors thank S. Girard and D. Eysseric for the technical design and support with the experimental campaign, as well as for their insightful discussions.}

\backsection[Funding]{The work of I.P.-M. has received funding from the European Union's Horizon Europe research and innovation programme under the Marie Sk\l{}odowska-Curie grant agreement no. 101063992. The work of D.R. has received funding from the Spanish Ministry of Science, Innovation and Universities and European Union's FEDER (PID2021-125812OB-C22). This work has also received funding from the Government of the Community of Madrid within the multi-annual agreement with Universidad Polit\'{e}cnica de Madrid through the Program of Excellence in Faculty (V-PRICIT line 3).}

\backsection[Declaration of interests]{The authors report no conflict of interest.}

\backsection[Data availability statement]{The data that support the findings of this study are available upon request.}

\backsection[Author ORCIDs]{I. Padilla-Montero, https://orcid.org/0000-0001-6643-4459; \\ D. Rodr\'{i}guez, https://orcid.org/0000-0002-1088-1927; \\ V. Jaunet, https://orcid.org/0000-0002-6272-1431; \\ P. Jordan, https://orcid.org/0000-0001-8576-5587.}

\appendix

\section{Comparison of mean flow solutions for different sets of constants of the Menter SST model}\label{appA}

\begin{figure}
\centerline{\includegraphics[width=0.99\textwidth]{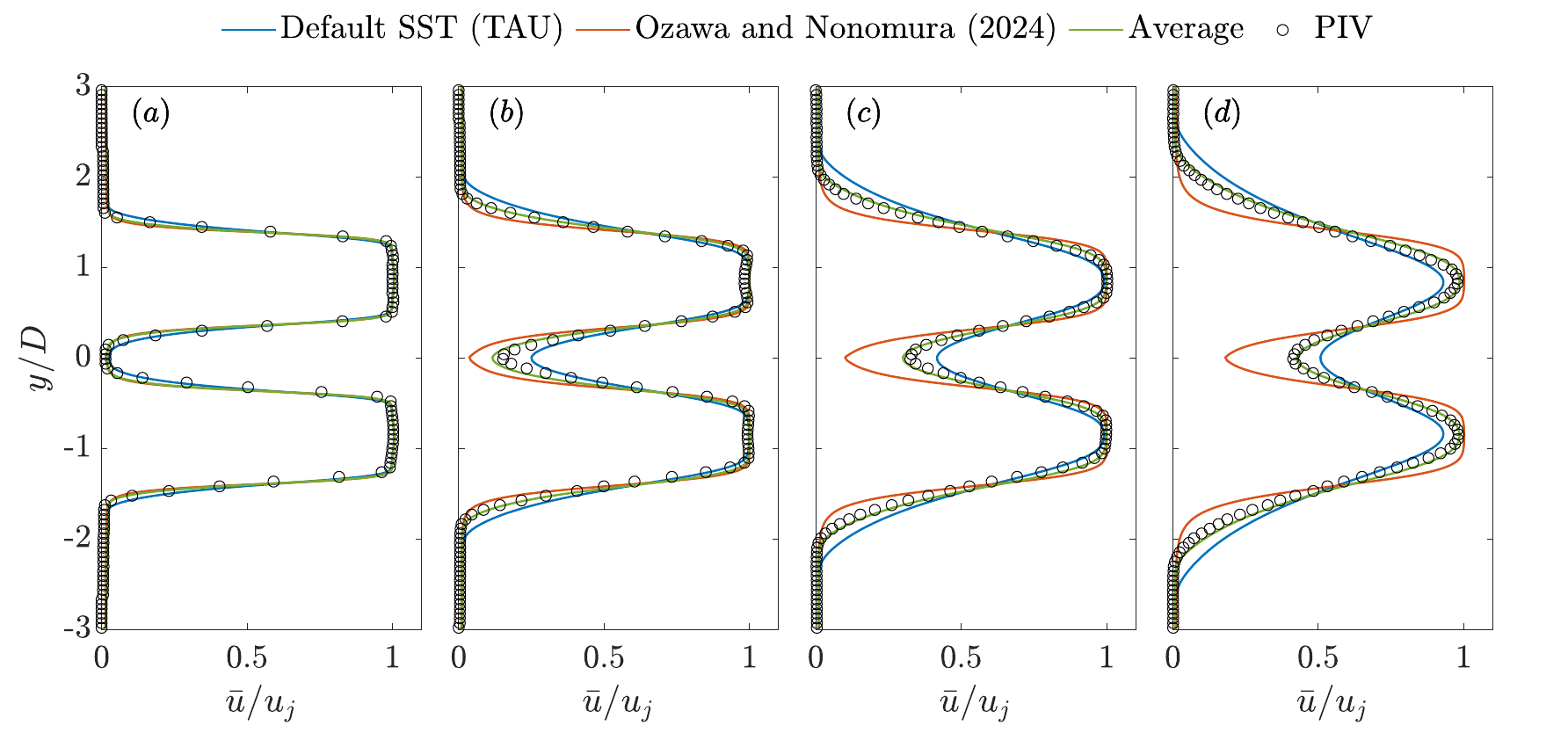}}
\caption{Comparison of mean streamwise velocity profiles along $y$ (and at $z = 0$) for four different streamwise locations, between the RANS solutions with different sets of SST model constants and the PIV mean flow ($s/D = 1.76$): $(a)$ $x/D = 2$; $(b)$ $x/D = 4$; $(c)$ $x/D = 6$; $(d)$ $x/D = 8$.}
\label{fig:RANS_vs_PIV_vs_x_all_SST}
\end{figure}

\begin{figure}
\centerline{\includegraphics[width=0.7\textwidth]{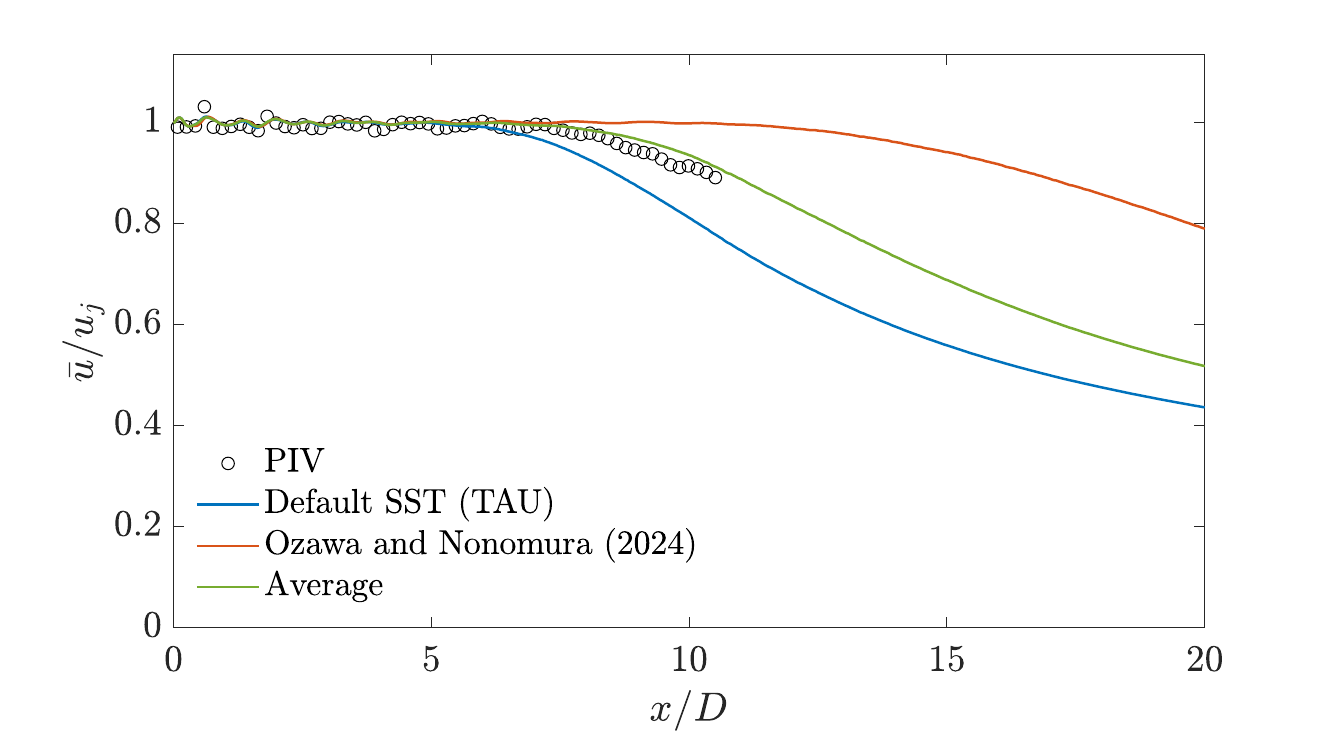}}
\caption{Comparison of mean streamwise velocity profiles along the top nozzle axis between the RANS solutions with different sets of SST model constants and the PIV mean flow for $s/D = 1.76$.}
\label{fig:RANS_vs_PIV_nzz_axis_all_SST}
\end{figure}

Figures~\ref{fig:RANS_vs_PIV_vs_x_all_SST} and~\ref{fig:RANS_vs_PIV_nzz_axis_all_SST} illustrate the mean flow comparison between the RANS solutions computed using three different sets of constants for the Menter SST model, namely, the default SST constants implemented in TAU, the optimized SST constants from~\citet{Ozawa2024} and the average between both, all against the PIV mean flow. The corresponding constants for each case are listed in table~\ref{tab:SST_constants_comp}. For the problem under study, the default set of constants is found to yield an overly diffusive solution (potential core is too short), while the use of~\citet{Ozawa2024}'s constants is found to yield a not enough diffusive solution (potential core is too long). For this reason, a set of constants corresponding to the average between both has been employed in this work, which is able to accurately reproduce the measured mean flow.

\begin{table}
  \begin{center}
\def~{\hphantom{0}}
  \begin{tabular}{lccccccccc}
      & $a$ & $\beta^*$ & $\kappa$ & $\beta_1$ & $\beta_2$ & $\sigma_{k1}$ & $\sigma_{k2}$ & $\sigma_{\omega 1}$ & $\sigma_{\omega 2}$ \\[3pt]
      Default (TAU solver) & 0.31 & 0.09 & 0.41 & 0.0752 & 0.0828 & 0.85 & 1 & 0.5 & 0.857\\
      \citet{Ozawa2024} & 0.434 & 0.08 & 0.41 & 0.07 & 0.102 & 0.474 & 0.574 & 0.393 & 0.634\\
      Average & 0.372 & 0.085 & 0.41 & 0.0726 & 0.0924 & 0.662 & 0.787 & 0.446 & 0.745\\
  \end{tabular}
  \caption{Different sets of values considered for the parameters of the Menter SST turbulence model. The same nomenclature as in \citet{Menter1994} is followed.}
  \label{tab:SST_constants_comp}
  \end{center}
\end{table}

\bibliographystyle{jfm}
\bibliography{referencesJFM_clean}

\end{document}